\documentclass{kapproc} 

\usepackage{epsfig}

\let\footnote\savefootnote

\let\footnoterule\savefootnoterule

\normallatexbib
\begin{document}

\articletitle[]{Cosmology, Inflation, and the Physics of Nothing}

\author{William H. Kinney}
\affil{Institute for Strings, Cosmology and Astroparticle Physics\\
       Columbia University\\
       550 W. 120th Street\\
       New York, NY 10027}
\email{kinney@physics.columbia.edu}

\begin{abstract}
These four lectures cover four topics in modern cosmology: the cosmological constant, the cosmic microwave background, inflation, and cosmology as a probe of physics at the Planck scale. The underlying theme is that cosmology gives us a unique window on the ``physics of nothing,'' or the quantum-mechanical properties of the vacuum. The theory of inflation postulates that vacuum energy, or something very much like it, was the dominant force shaping the evolution of the very early universe. Recent astrophysical observations indicate that vacuum energy, or something very much like it, is also the dominant component of the universe today. Therefore cosmology gives us a way to study an important piece of particle physics inaccessible to accelerators. The lectures are oriented toward graduate students with only a passing familiarity with general relativity and knowledge of basic quantum field theory. 
\end{abstract}

\section{Introduction}

Cosmology is undergoing an explosive burst of activity, fueled both by new, accurate astrophysical
data and by innovative theoretical developments. Cosmological parameters such as the total
density of the universe and the rate of cosmological expansion are being precisely measured for
the first time, and a consistent standard picture of the universe is beginning to emerge. This
is exciting, but why talk about astrophysics at a school for particle physicists? The answer
is that over the past twenty years or so, it has become evident that the the story of the
universe is really a story of fundamental physics.  I will argue that not only should particle 
physicists care about cosmology, but you should care  {\em a lot}. Recent developments in 
cosmology indicate that
it will be possible to use astrophysics to perform tests of fundamental theory 
inaccessible to particle accelerators, namely the physics of the vacuum itself. This has proven
to be a surprise to cosmologists: the old picture of a universe filled only with
matter and light have given way to a picture of a universe whose history is largely written
in terms of the quantum-mechanical properties of empty space. It is currently believed that
the universe today is dominated by the energy of vacuum, about 70\% by weight. In addition,
the idea of inflation postulates that the universe at the earliest times in its history was
also dominated by vacuum energy, which introduces the intriguing possibility that
all structure in the universe, from superclusters to planets, had a quantum-mechanical
origin in the earliest moments of the universe. Furthermore, these ideas are not idle
theorizing, but are predictive and subject to meaningful experimental test. Cosmological
observations are providing several surprising challenges to fundamental theory. 

These lectures are organized as follows. Section \ref{secintro} provides an introduction
to basic cosmology and a description of the surprising recent discovery of the accelerating 
universe. Section \ref{seccmb} discusses the physics of the cosmic microwave background
(CMB), one of the most useful observational tools in modern cosmology. Section 
\ref{secinflation} discusses some unresolved problems in standard Big-Bang cosmology, and
introduces the idea of inflation as a solution to those problems. Section \ref{sectransplanck}
discusses the intriguing (and somewhat speculative) idea of using inflation as a 
``microscope'' to illuminate physics at the very highest energy scales, where effects
from quantum gravity are likely to be important. These lectures are geared toward graduate
students who are familiar with special relativity and quantum mechanics, and who have
at least been introduced to general relativity and quantum field theory. There are many
things I will not talk about, such as dark matter and structure formation, which are 
interesting but do not touch directly on the main theme of the ``physics of nothing.'' I omit
many details, but I provide references to texts and review articles where possible.

\section{Resurrecting Einstein's greatest blunder.}
\label{secintro}

\subsection{Cosmology for beginners}

All of modern cosmology stems essentially from an application of the Copernican principle:
we are not at the center of the universe. In fact, today we take Copernicus' idea one
step further and assert the ``cosmological principle'': {\em nobody} is at the center of
the universe. The cosmos, viewed from any point, looks the same as when viewed from any
other point. This, like other symmetry principles more directly familiar to particle
physicists, turns out to be an immensely powerful idea. In particular, it leads to the
apparently inescapable conclusion that the universe has a finite age. There was a
beginning of time. 

We wish to express the cosmological principle mathematically, as a symmetry. To do this,
and to understand the rest of these lectures, we need to talk about metric tensors and
General Relativity, at least briefly. A {\em metric} on a space is simply a generalization
of Pythagoras' theorem for the distance $ds$ between two points separated by distances
$d{\bf x} = (dx,dy,dz)$,
\begin{equation}
ds^2 = \left\vert d{\bf x}\right\vert^2 = dx^2 + dy^2 + dz^2.
\end{equation}
We can write this as a matrix equation,
\begin{equation}
ds^2 = \sum_{i,j = 1, 3} \eta_{i j} dx^i dx^j,
\end{equation}
where $\eta_{i j}$ is just the unit matrix,
\begin{equation}
\eta_{i j} = \left(\matrix{1&0&0\cr0&1&0\cr0&0&1}\right).
\end{equation}
The matrix $\eta_{i j}$ is referred to as the {\em metric} of the space, in this case
a three-dimensional Euclidean space. One can define other, non-Euclidean spaces by specifying
a different metric. A familiar one is the four-dimensional ``Minkowski'' space of special
relativity, where the proper distance between two points in spacetime is given by
\begin{equation}
ds^2 = dt^2 - d{\bf x}^2,
\end{equation}
corresponding to a metric tensor with indices $\mu,\nu = 0,\ldots,3$:
\begin{equation}
\label{eqminkowskimetric}
\eta_{\mu \nu} = \left(\matrix{1&0&0&0\cr0&-1&0&0\cr0&0&-1&0\cr0&0&0&-1}\right).
\end{equation}
In a Minkowski space, photons travel on null paths, or {\em geodesics}, $ds^2 = 0$, and massive 
particles travel on {\em timelike} geodesics, $ds^2 > 0$. Note that in both of the examples
given above, the metric is time-independent, describing a static space. In General Relativity,
the metric becomes a dynamic object, and can in general depend on time and space. The fundamental
equation of general relativity is the Einstein field equation,
\begin{equation}
\label{eqEFE}
G_{\mu \nu} = 8 \pi G T_{\mu \nu},
\end{equation}
where $T_{\mu \nu}$ is a {\em stress energy} tensor describing the distribution of mass in space,
G is Newton's gravitational constant and the Einstein Tensor $G_{\mu \nu}$ is a complicated 
function of the metric and its first
and second derivatives. This should be familiar to anyone who has taken a course in 
electromagnetism, since we can write Maxwell's equations in matrix form as
\begin{equation}
\label{eqMAX}
\partial_\nu F^{\mu \nu} = {4 \pi \over c} J^{\mu},
\end{equation}
where $F^{\mu \nu}$ is the field tensor and $J^{\mu}$ is the current. Here we use the standard
convention that we sum over the repeated indices of four-dimensional spacetime $\nu = 0, 3$. 
Note the similarity 
between Eq. (\ref{eqEFE}) and Eq. (\ref{eqMAX}). The similarity is more than formal: both have a
charge on the right hand side acting as a source for a field on the left hand side. In the
case of Maxwell's equations, the source is electric charge and the field is the electromagnetic
field. In the case of Einstein's equations, the source is mass/energy, and the field is 
the shape of the spacetime, or the metric. An additional feature of the Einstein field equation
is that it is {\em much} more complicated than Maxwell's equations: Eq. (\ref{eqEFE}) represents
six independent nonlinear partial differential equations of ten functions, the components
of the (symmetric) metric tensor $g_{\mu \nu}(t, {\bf x})$. (The other four degrees of 
freedom are accounted for by invariance under transformations among the four coordinates.)

Clearly, finding a general solution to a set of equations as complex as the Einstein field 
equations is a hopeless task. Therefore, we do what any good physicist does when faced with
an impossible problem: we introduce a symmetry to make the problem simpler. The three simplest
symmetries we can apply to the Einstein field equations are: (1) vacuum, (2) spherical symmetry,
and (3) homogeneity and isotropy. Each of these symmetries is useful (and should be familiar).
The assumption of vacuum is just the case where there's no matter at all:
\begin{equation}
T_{\mu \nu} = 0.
\end{equation}
In this case, the Einstein field equation reduces to a wave equation, and the solution 
is gravitational radiation. If we assume that the matter distribution $T_{\mu\nu}$ has
spherical symmetry, the solution to the Einstein field equations is the Schwarzschild
solution describing a black hole. The third case, homogeneity and isotropy, is the one
we will concern ourselves with in more detail here \cite{weinberg}. By homogeneity, we mean that the 
universe is invariant under spatial translations, and by isotropy we mean that the universe
is invariant under rotations. (A universe that is isotropic everywhere is necessarily 
homogeneous, but a homogeneous universe need not be isotropic: imagine a homogeneous space
filled with a uniform electric field!) We will model the contents of  the universe as a 
perfect fluid with density $\rho$ and pressure $p$, for which the stress-energy tensor is
\begin{equation}
\label{eqfluidse}
T^{\mu}{}_{\!\nu} = \left(\matrix{\rho&0&0&0\cr0&-p&0&0\cr0&0&-p&0\cr0&0&0&-p}\right).
\end{equation}
While this is certainly a poor description of the contents of the universe on small scales,
such as the size of people or planets or even galaxies, it is an excellent approximation
if we average over extremely large scales in the universe, for which the matter is known
observationally to be very smoothly distributed. If the matter in the universe is homogeneous
and isotropic, then the metric tensor must also obey the symmetry. The most general line element
consistent with homogeneity and isotropy is 
\begin{equation}
\label{eqRWmetric}
ds^2 = dt^2 - a^2(t)d{\bf x}^2,
\end{equation}
where the {\em scale factor} $a(t)$ contains all the dynamics of the universe, and the
vector product $d{\bf x}^2$ contains the geometry of the space, which can be either Euclidean
($d{\bf x}^2 = dx^2 + dy^2 + dz^2$) or positively or negatively curved. The metric tensor
for the Euclidean case is particularly simple,
\begin{equation}
g_{\mu \nu} = \left(\matrix{1&0&0&0\cr0&-a(t)&0&0\cr0&0&-a(t)&0\cr0&0&0&-a(t)}\right),
\end{equation}
which can be compared to the Minkowski metric (\ref{eqminkowskimetric}). In this
{\em Friedmann-Robertson-Walker} (FRW) space, spatial distances are multiplied by a 
dynamical factor $a(t)$ that describes the expansion (or contraction) of the spacetime.
With the general metric (\ref{eqRWmetric}), the Einstein field equations take on a 
particularly simple form,
\begin{equation}
\label{eqRW}
\left({\dot a \over a}\right)^2 = {8 \pi G \over 3} \rho - {k \over a^2},
\end{equation}
where $k$ is a constant that describes the curvature of the space: $k = 0$ (flat), or
$k = \pm 1$ (positive or negative curvature). This is known as the {\em Friedmann equation}.
In addition, we have a second-order equation
\begin{equation}
\label{eqraychaud}
{\ddot a \over a} = - {4 \pi G \over 3} \left(\rho + 3 p\right).
\end{equation}
Note that the second derivative of the scale factor depends on the equation of state
of the fluid.  The equation of state is frequently given by a parameter $w$, or
$p = w \rho$. Note that for any fluid with positive pressure, $w > 0$, the expansion of
the universe is gradually decelerating, $\ddot a < 0$: the mutual gravitational attraction 
of the matter in the universe slows the expansion. This characteristic will be central
to the discussion that follows. 

\subsection{Einstein's ``greatest blunder''}

General relativity combined with homogeneity and isotropy leads to a startling conclusion:
spacetime is dynamic. The universe is not static, but is bound to be either expanding or
contracting. In the early 1900's, Einstein applied general relativity to the homogeneous and
isotropic case, and upon seeing the consequences, decided that the answer had to be wrong.
Since the universe was obviously static, the equations had to be fixed. Einstein's method
for fixing the equations involved the evolution of the density $\rho$ with expansion. 
Returning to our analogy between General Relativity and electromagnetism, we remember that
Maxwell's equations (\ref{eqMAX}) imply the conservation of charge,
\begin{equation}
\partial_\mu J^{\mu} = 0,
\end{equation}
or, in vector notation,
\begin{equation}
{\partial  \rho \over \partial t} + \nabla \cdot {\bf J} = 0.
\end{equation}
The general relativistic equivalent to charge conservation is stress-energy conservation,\footnote{As in the case of Maxwell's equations, stress-energy conservation is not  independent of the Einstein field equations, but is a consequence of the Bianchi identities. This can be seen by noting that Eqs. (\ref{eqRW}) and (\ref{eqraychaud})
taken together imply Eq. (\ref{eqcontinuity}).}
\begin{equation}
\label{eqsecons}
D_{\mu} T^{\mu \nu} = 0.
\end{equation}
For a homogeneous fluid with the stress-energy given by Eq. (\ref{eqfluidse}), stress-energy 
conservation takes the form of the {\em continuity equation},
\begin{equation}
\label{eqcontinuity}
{d \rho \over d t} + 3 H \left(\rho + p\right) = 0,
\end{equation}
where $H = \dot a / a$ is the Hubble parameter from Eq. (\ref{eqRW}). This equation relates
the evolution of the energy density to its equation of state $p = w \rho$. Suppose we have
a box whose dimensions are expanding along with the universe, so that the volume of the
box is proportional to the cube of the scale factor, $V \propto a^3$, and we fill it with
some kind of matter or radiation. For example, ordinary matter in a box of volume $V$ has 
an energy density inversely proportional to the  volume of the box, 
$\rho \propto V^{-1} \propto a^{-3}$.
It is straightforward to show using the continuity equation that this corresponds to zero 
pressure, $p = 0$. Relativistic particles such as photons have energy density that goes as
$\rho \propto V^{-4/3} \propto a^{-4}$, which corresponds to equation of state $p = \rho/3$.

Einstein noticed that if we take the stress-energy $T_{\mu \nu}$ and add a constant $\Lambda$, 
the conservation equation (\ref{eqsecons}) is unchanged:
\begin{equation}
D_{\mu} T^{\mu \nu} = D_{\mu} \left(T^{\mu \nu} + \Lambda g^{\mu \nu}\right) = 0.
\end{equation}
In our analogy with electromagnetism, this is like adding a constant to the 
electromagnetic potential, $V'(x) = V(x) + \Lambda$. The constant $\Lambda$ does not
affect local dynamics in any way, but it does affect the cosmology. Since adding this
constant adds a constant energy density to the universe, the continuity equation tells
us that this is equivalent to a fluid with {\em negative} pressure, 
$p_{\Lambda} = -\rho_{\Lambda}$. Einstein chose  $\Lambda$ to give a closed, static 
universe as follows \cite{einsteinstatic}. Take the energy density to consist of matter
\begin{eqnarray}
\rho_{\rm M} &&= {k \over 4 \pi G a^2}\cr
p_{\rm M} && = 0,
\end{eqnarray}
and cosmological constant
\begin{eqnarray}
\rho_{\Lambda} &&= {k \over 8 \pi G a^2}\cr
p_{\rm \Lambda} && =  - \rho_{\Lambda}.
\end{eqnarray}
It is a simple matter to use the Friedmann equation to show that this combination of matter
and cosmological constant leads to a static universe $\dot a = \ddot a = 0$. In order for
the energy densities to be positive, the universe must be closed, $k = +1$. Einstein was
able to add a kludge to get the answer he wanted.

Things sometimes happen in science with uncanny timing. In the 1920's, an astronomer named
Edwin Hubble undertook a project to measure the distances to the spiral ``nebulae'' as 
they had been known, using the 100-inch Mount Wilson telescope. Hubble's method involved
using Cepheid variables, named after the star Delta Cephei, the best known member of the
class.\footnote{Delta Cephei is not, however the nearest Cepheid. That honor goes to Polaris,
the north star \cite{hipparcos}.} Cepheid variables have the useful property that the period of their 
variation, usually 10-100 days, is correlated to their absolute brightness. Therefore, by
measuring the apparent brightness and the period of a distant Cepheid, one can determine
its absolute brightness and therefore its distance. Hubble applied this method to a number
of nearby galaxies, and determined that almost all of them were receding from the earth. 
Moreover, the more distant the galaxy was, the faster it was receding, according to a
roughly linear relation:
\begin{equation}
\label{eqHubblelaw}
v = H_0 d.
\end{equation}
This is the famous Hubble Law, and the constant $H_0$ is known as Hubble's constant. Hubble's
original value for the constant was something like $500\ {\rm km/sec/Mpc}$, where one megaparsec
(${\rm Mpc}$) is a bit over 3 million light years.\footnote{The {\em parsec} is an archaic 
astronomical unit corresponding to one second of arc of parallax measured from opposite sides
of the earth's orbit: $1\ {\rm pc} = 3.26\ {\rm ly}$.} 
This implied an age for the universe of about a
billion years, and contradicted known geological estimates for the age
of the earth. Cosmology had its first ``age problem'': the universe can't be younger than
the things in it! Later it was realized that Hubble had failed to account for two distinct
types of Cepheids, and once this discrepancy was taken into account, the Hubble constant fell
to well under $100\ {\rm km/s/Mpc}$. The current best estimate, determined using the Hubble space
telescope to resolve Cepheids in galaxies at unprecedented distances, is 
$H_0 = 71 \pm 6\ {\rm km/s/Mpc}$ \cite{Freedman99}. In any case, the Hubble law is exactly
what one would expect from the Friedmann equation. The expansion of the universe
predicted (and rejected) by Einstein had been observationally detected, only a few years
after the development of General Relativity. Einstein is said to have later referred to the introduction of the cosmological constant as his ``greatest blunder'' (a quote which may be apocryphal: see p. 9 of the review by Padmanabhan \cite{padman02}).

The expansion of the universe leads to a number of interesting things. One is the
cosmological redshift of photons. The usual way to see this is that from the Hubble law, 
distant objects appear to be receding at a velocity $v = H_0 d$, which means that photons 
emitted from the body are redshifted due to the recession velocity of the source. There is
another way to look at the same effect: because of the expansion of space, the wavelength
of a photon increases with the scale factor:
\begin{equation}
\lambda \propto a(t),
\end{equation}
so that as the universe expands, a photon propagating in the space gets shifted to longer
and longer wavelengths. The redshift $z$ of a photon is then given by the ratio of the scale
factor today to the scale factor when the photon was emitted:
\begin{equation}
1 + z = {a(t_0) \over a(t_{\rm em})}.
\end{equation}
Here we have introduced commonly used  the convention that a subscript $0$ (e.g., $t_0$ or $H_0$) 
indicates the value of a quantity {\em today}. This redshifting due to expansion applies
to particles other than photons as well. For some massive body moving relative to the 
expansion with some momentum $p$, the momentum also ``redshifts'':
\begin{equation}
p \propto {1 \over a(t)}.
\end{equation}
We then have the remarkable result that freely moving bodies in an expanding universe eventually
come to rest relative to the expanding coordinate system, the so-called {\em comoving} frame. 
The expansion of the universe creates a kind of dynamical friction for everything moving
in it. For this reason, it will often be convenient to define comoving variables, which
have the effect of expansion factored out. For example, the physical distance between two
points in the expanding space is proportional to $a(t)$. We define the comoving distance between 
two points to be a constant in time:
\begin{equation}
x_{\rm com} = x_{\rm phys} / a(t) = {\rm const.}
\end{equation}
Similarly, we define the comoving wavelength of a photon as
\begin{equation}
\lambda_{\rm com} = \lambda_{\rm phys} / a(t),
\end{equation}
and comoving momenta are defined as:
\begin{equation}
p_{\rm com} \equiv a(t) p_{\rm phys}.
\end{equation}
This energy loss with expansion has a predictable effect on systems in thermal equilibrium.
If we take some bunch of particles (say, photons with a black-body distribution) in thermal 
equilibrium with temperature $T$, the momenta of all these particles will decrease linearly
with expansion, and the system will cool.\footnote{It is not hard to convince yourself that
a system that starts out as a blackbody stays a blackbody with expansion.} For a gas in 
thermal equilibrium, the temperature is in fact inversely proportional to the scale factor:
\begin{equation}
T \propto {1 \over a(t)}.
\end{equation}
The current temperature of the universe is about $3\ {\rm K}$. Since it has been cooling with
expansion, we reach the conclusion that the early universe must have been at a much higher
temperature. This is the ``Hot Big Bang'' picture: a hot, thermal equilibrium universe  
expanding and cooling with time. One thing to note is that, although the universe goes to
infinite density and infinite temperature at the Big Bang singularity, it does {\em not}
necessarily go to zero size. A flat universe, for example is infinite in spatial extent
an infinitesimal amount of time after the Big Bang, which happens {\em everywhere} in the
infinite space simultaneously! The observable universe, as measured by the horizon size, 
goes to zero  size at $t = 0$, but the observable universe represents only a tiny patch of 
the total space.

\subsection{Critical density and the return of the age problem}

One of the things that cosmologists most want to measure accurately is the total density
$\rho$ of the universe. This is most often expressed in units of the density needed to
make the universe flat, or $k = 0$. Taking the Friedmann equation for a $k = 0$ universe,
\begin{equation}
H^2 = \left({\dot a \over a}\right)^2 = {8 \pi G \over 3} \rho,
\end{equation}
we can define a critical density $\rho_c$,
\begin{equation}
\rho_c \equiv {3 H_0^2 \over 8 \pi G},
\end{equation}
which tells us, for a given value of the Hubble constant $H_0$, the energy density of
a Euclidean FRW space. If the energy density is greater than critical, $\rho > \rho_c$,
the universe is closed and has a positive curvature ($k = +1$). In this case, the
universe also has a finite lifetime, eventually collapsing back on itself in a ``big
crunch''. If $\rho < \rho_c$, the universe is open, with negative curvature, and has an
infinite lifetime. This is usually expressed in terms of the density parameter $\Omega$,
\begin{eqnarray}
&&< 1:\ {\rm Open}\cr
\Omega \equiv {\rho \over \rho_c} &&= 1:\ {\rm Flat}\cr
&&>1:\ {\rm Closed}.
\end{eqnarray}
There has long been a debate between theorists and observers as to what the
value of $\Omega$ is in the real universe. Theorists have steadfastly maintained that
the only sensible value for $\Omega$ is unity, $\Omega = 1$. This prejudice was further
strengthened by the development of the theory of inflation, which solves several 
cosmological puzzles (see Secs. \ref{secflatnessproblem} and \ref{sechorizonproblem}) and
in fact {\em predicts} that $\Omega$ will be exponentially close to unity. Observers,
however, have made attempts to measure $\Omega$ using a variety of methods, including
measuring galactic rotation curves, the velocities of galaxies orbiting in clusters,
X-ray measurements of galaxy clusters, the velocities and spatial distribution of galaxies 
on large scales, and gravitational lensing. These measurements have repeatedly pointed
to a value of $\Omega$ inconsistent with a flat cosmology, with $\Omega = 0.2-0.3$ being
a much better fit, indicating an open, negatively curved universe. Until a few years
ago, theorists have resorted to cheerfully ignoring the data, taking it almost on faith that 
$\Omega = 0.7$ in extra stuff would turn up sooner or later. The theorists were right: new
observations of the cosmic microwave background definitively favor a flat universe, $\Omega = 1$.
Unsurprisingly, the observationalists were also right: only about $1/3$ of this density
appears to be in the form of ordinary matter.  

The first hint that something strange was up with the standard cosmology came from
measurements of the colors of stars in globular clusters. Globular clusters are small,
dense groups of $10^{5}$ - $10^{6}$ stars which orbit in the halos of most galaxies and
are among the oldest objects in the universe. Their ages are determined by observation
of stellar populations and models of stellar evolution, and some globular clusters are
known to be at least 12 billion years old \cite{globularages}, implying that the universe itself must
be at least 12 billion years old. But consider a flat universe ($\Omega = 1$) filled
with pressureless matter, $\rho \propto a^{-3}$ and $p = 0$. It is straightforward to
solve the Friedmann equation (\ref{eqRW}) with $k = 0$ to show that
\begin{equation}
a\left(t\right) \propto t^{2/3}.
\end{equation}
The Hubble parameter is then given by
\begin{equation}
H = {\dot a \over a} = {2 \over 3} t^{-1}.
\end{equation}
We therefore have a simple expression for the age of the universe $t_0$ in terms of the
measured Hubble constant $H_0$, 
\begin{equation}
t_0 = {2 \over 3} H_0^{-1}.
\end{equation}
The fact that the universe has a finite age introduces the concept of a {\em horizon}: this
is just how far out in space we are capable of seeing at any given time. This distance is
finite because photons have only traveled a finite distance since the beginning of the 
universe. Just as in special relativity, photons travel on paths with proper length 
$ds^2 = dt^2 - a^2 d{\bf x}^2 = 0$, so that we can write the physical distance a photon has 
traveled since the Big Bang, or the {\em horizon size}, as
\begin{equation}
d_{\rm H} = a(t_0) \int_{0}^{t_0}{dt \over a(t)}.
\end{equation}
(This is in units where the speed of light is set to $c = 1$.) For example, in a flat,
matter-dominated universe, $a(t) \propto t^{2/3}$, so that the horizon size is
\begin{equation}
d_{\rm H} = t_0^{2/3} \int_{0}^{t0}{t^{-2/3} d t} = 3 t_0 = 2 H_0^{-1}.
\end{equation}
This form for the horizon distance is true in general: the distance a photon travels in 
time $t$ is always about $d \sim t$: effects from expansion simply add a numerical factor
out front. We will frequently ignore this, and approximate
\begin{equation}
d_{\rm H} \sim t_0 \sim H_0^{-1}.
\end{equation}
Measured values of $H_0$ are quoted in a very strange unit of time, a km/s/Mpc, but it is
a simple matter to calculate the dimensionless factor using
$1\ {\rm Mpc} \simeq 3 \times 10^{19}\ {\rm km}$,
so that the age of a flat, matter-dominated universe with $H_0 = 71 \pm 6\ {\rm km/s/Mpc}$ is
\begin{equation}
t_0 = 8.9^{+ 0.9}_{- 0.7} \times 10^{9}\ {\rm years}.
\end{equation}
A flat, matter-dominated universe would be younger than the things in it! Something is
evidently wrong -- either the estimates of globular cluster ages are too big, the 
measurement of the Hubble constant from from the HST is incorrect, the universe is 
not flat, or the universe is not matter dominated. 

\begin{figure}[ht]
\vskip0.5cm
\centerline{\epsfig{file=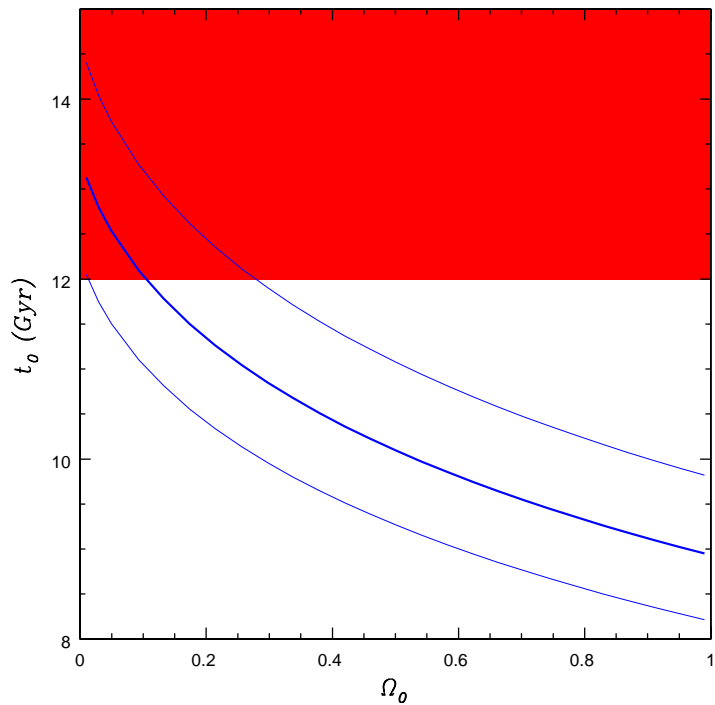, width=10.0cm}}
\caption{Age of the universe as a function of $\Omega_0$ for a matter-dominated universe. The
blue lines show the age $t_0$ consistent with the HST key project value 
$H_0 = 72 \pm 6\ {\rm km/s/Mpc}$. The red area is the region consistent with globular 
cluster ages $t_0 > 12\ {\rm Gyr}$.}
\label{figage}
\end{figure}

We will take for granted that the measurement of the Hubble constant is correct, and that
the models of stellar structure are good enough to produce a reliable estimate of globular
cluster ages (as they appear to be), and focus on the last two possibilities. An open
universe, $\Omega_0 < 1$, might solve the age problem. Figure \ref{figage} shows the age of
the universe consistent with the HST Key Project value for $H_0$ as a function of the
density parameter $\Omega_0$.
We see that the age determined from $H_0$ is consistent with globular clusters as old as
12 billion years only for values of $\Omega_0$ less than $0.3$ or so. However, as we will see 
in Sec.
\ref{seccmb}, recent measurements of the cosmic microwave background strongly indicate 
that we indeed live in a flat ($\Omega = 1$) universe. So while a low-density universe might
provide a marginal solution to the age problem, it would conflict with the CMB. We
therefore, perhaps reluctantly, are forced to consider that the universe might not be
matter dominated. In the next section we will take a detour into quantum field theory 
seemingly unrelated to these cosmological issues. By the time we are finished, however,
we will have in hand a different, and provocative, solution to the age problem consistent
with a flat universe. 

\subsection{The vacuum in quantum field theory}
\label{secqftbasics}

In this section, we will discuss something that at first glance appears to be entirely
unrelated to cosmology: the vacuum in quantum field theory. We will see later, however,
that it will in fact be crucially important to cosmology. Let us start with basic
quantum mechanics, in the form of the simple harmonic oscillator, with Hamiltonian
\begin{equation}
H = \hbar \omega\left(\hat a^{\dagger} \hat a + {1\over 2}\right),
\end{equation}
where $\hat a$ and $\hat a^{\dagger}$ are the lowering and raising operators, respectively, with
commutation relation
\begin{equation}
\left[\hat a,\hat a^{\dagger}\right] = 1.
\end{equation}
This leads to the familiar ladder of energy eigenstates $\left\vert n \right\rangle$,
\begin{equation}
H \left\vert n \right\rangle = \hbar \omega \left(n + {1 \over 2}\right) \left\vert n \right\rangle
= E_n \left\vert n \right\rangle.
\end{equation}
The simple harmonic oscillator is pretty much the only problem that physicists know how
to solve. Applying the old rule that if all you have is a hammer, everything looks like
a nail, we construct a description of quantum fields by placing an infinite number of
harmonic oscillators at every point,
\begin{equation}
H = \int^{\infty}_{-\infty}{d^3 k \left[\hbar \omega_k \left({\hat a_{\bf k}}^{\dagger} {\hat a_{\bf k}} + {1 \over 2}\right)\right]},
\end{equation}
where the operators ${\hat a_{\bf k}}$ and ${\hat a_{\bf k}}^{\dagger}$ have the commutation relation
\begin{equation}
\left[{\hat a_{\bf k}}, \hat a_{\bf k'}^{\dagger}\right] = \delta^3\left({\bf k} - {\bf k'}\right).
\end{equation}
Here we make the identification that  ${\bf k}$ is the momentum of a particle, and $\omega_{k}$
is the energy of the particle, 
\begin{equation}
\omega_k^2 - \left\vert {\bf k}\right\vert^2 = m^2
\end{equation}
for a particle of mass $m$. Taking $m = 0$ gives the familiar dispersion relation for massless
particles like photons. Like the state kets $\left\vert n \right\rangle$ for the harmonic 
oscillator, each momentum vector $\bf k$ has an independent ladder of states, with the associated
quantum numbers, $\left\vert n_{\bf k}, \ldots, n_{\bf k'}\right\rangle$. The raising and lowering
operators are now interpreted as {\em creation} and {\em annihilation} operators, turning a 
ket with $n$ particles into a ket with $n + 1$ particles, and vice-versa:
\begin{equation}
\left\vert n_{\bf k} = 1\right\rangle = {\hat a_{\bf k}}^{\dagger} \left\vert 0 \right\rangle,
\end{equation}
and we call the ground state $\left\vert 0 \right\rangle$ the {\em vacuum}, or zero-particle
state. But there is one small problem: just like the ground state of a single harmonic
oscillator has a nonzero energy $E_0 = (1/2) \hbar \omega$, the vacuum state of the quantum
field also has an energy,
\begin{eqnarray}
\label{eqdivergentvacuum}
H \left\vert 0\right\rangle &=& \int^{\infty}_{-\infty}{d^3 k \left[\hbar \omega_k \left({\hat a_{\bf k}}^{\dagger} {\hat a_{\bf k}} + {1 \over 2}\right)\right]} \left\vert 0 \right\rangle\cr
&=& \left[\int^{\infty}_{-\infty}{d^3 k \left(\hbar \omega_k / 2\right)}\right] \left\vert 0 \right\rangle\cr
&=& \infty.
\end{eqnarray}
The ground state energy diverges! The solution to this apparent paradox is that we expect
quantum field theory to break down at very high energy. We therefore introduce  a cutoff on
the momentum ${\bf k}$ at high energy, so that the integral in Eq. (\ref{eqdivergentvacuum})
becomes finite. A reasonable physical scale for the cutoff is the scale at which we expect
quantum gravitational effects to become relevant, the {\em Planck scale} $m_{\rm Pl}$:
\begin{equation}
H\left\vert 0 \right\rangle \sim m_{\rm Pl} \sim 10^{19}\ {\rm GeV}.
\end{equation}
Therefore we expect the vacuum everywhere to have a constant energy density, given in units
where $\hbar = c = 1$ as
\begin{equation}
\rho \sim m_{\rm Pl}^4 \sim 10^{93}\ {\rm g/cm^3}.
\end{equation}
But we have already met up with an energy density that is constant everywhere in space: Einstein's
cosmological constant, $\rho_{\Lambda} = {\rm const.}$,~$p_{\Lambda} = - \rho_{\Lambda}$. 
However, the cosmological constant we expect from quantum field theory is more than a hundred
twenty orders of magnitude too big. In order for $\rho_\Lambda$ to be less than the critical
density, $\Omega_{\Lambda} < 1$, we must have 
$\rho_{\Lambda} < 10^{-120} m_{\rm Pl}^4$! How can we explain this discrepancy? 
Nobody knows. 

\subsection{Vacuum energy in cosmology}

So what does this have to do with cosmology? The interesting fact about vacuum energy is that
it results in accelerated expansion of the universe. From Eq. (\ref{eqraychaud}), we can
write the acceleration $\ddot a$ in terms of the equation of state $p = w \rho$ of the matter
in the universe,
\begin{equation}
{\ddot a \over a} \propto - \left(1 + 3 w\right) \rho.
\end{equation}
For ordinary matter such as pressureless dust $w = 0$ or radiation $w = 1/3$, we see that the
gravitational attraction of all the stuff in the universe makes the expansion slow down with
time, $\ddot a < 0$. But we have seen that a cosmological constant has the odd property of
negative pressure, $w = -1$, so that a universe dominated by vacuum energy actually expands
faster and faster with time, $\ddot a > 0$. It is easy to see that accelerating expansion
helps with the age problem: for a standard matter-dominated universe, a larger Hubble constant
means a {\em younger universe}, $t_0 \propto H_0^{-1}$. But if the expansion of the universe
is accelerating, this means that $H$ grows with time. For a given age $t_0$, acceleration means
that the Hubble constant we measure will be larger in an accelerating cosmology than in a
decelerating one, so we can have our cake and eat it too: an old universe and a high Hubble
constant! This also resolves the old dispute between the observers and the theorists. Astronomers
measuring the density of the universe use local dynamical measurements such as the orbital
velocities of galaxies in a cluster. These measurements are insensitive to a cosmological
constant and only measure the {\em matter} density $\rho_{\rm M}$ of the universe. However, 
geometrical tests like the cosmic microwave background which we will discuss in the Sec. 
\ref{seccmb} are sensitive to the {\em total} energy density $\rho_{\rm M} + \rho_{\rm \Lambda}$.
If we take the observational value for the matter density $\Omega_{\rm M} = 0.2-0.3$ and
make up the difference with a cosmological constant, $\Omega_{\Lambda} = 0.7-0.8$, we arrive
at an age for the universe in excess of $13\ {\rm Gyr}$, perfectly consistent with the
globular cluster data. 

\begin{figure}[ht]
\vskip0.5cm
\centerline{\epsfig{file=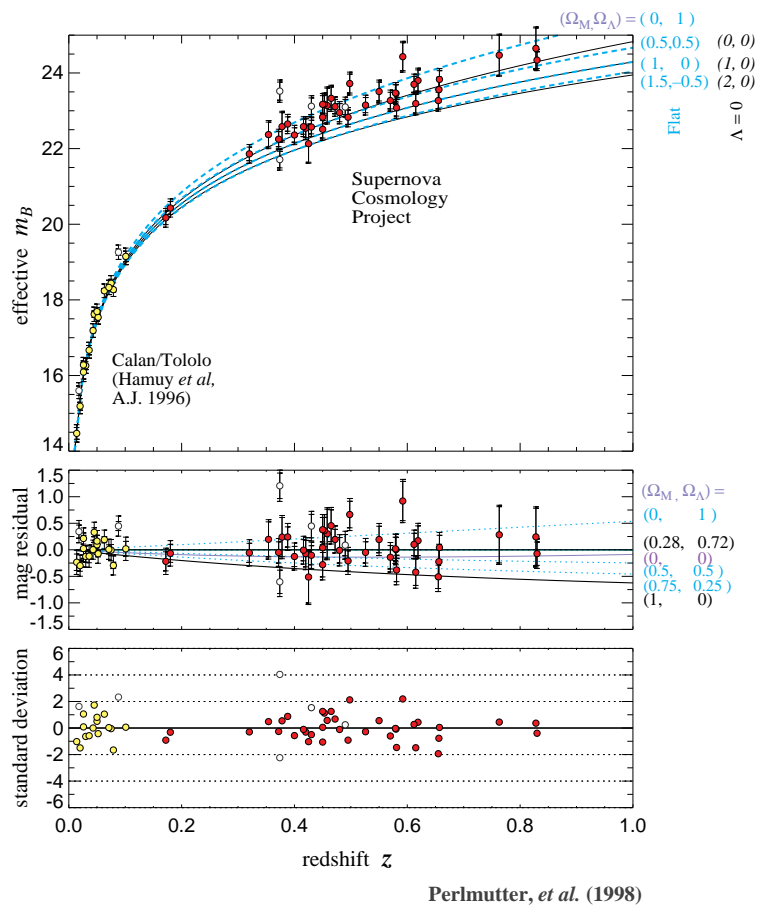, width=10.0cm}}
\caption{Data from the Supernova Cosmology project \cite{SCPHubblediag}. Dimmer objects are 
higher vertically
on the plot. The horizontal axis is redshift. The curves represent different choices of
$\Omega_{\rm M}$ and $\Omega_\Lambda$. A cosmology with $\Omega_{\rm M} = 1$ and 
$\Omega_\Lambda = 0$ is ruled out to 99\% confidence, while a universe with 
$\Omega_{\rm M} = 0.3$ and $\Omega_{\Lambda} = 0.7$ is a good fit to the data. }
\label{figSCP}
\end{figure}

In the 1980s and 1990s, there were some researchers making the argument based on the age problem
alone that we needed a cosmological constant \cite{krausslambda}. There were also some observational indications favoring a cosmological constant \cite{Jackson96}.
But the case was hardly
compelling, given that the CMB results indicating a flat universe had not yet been measured,
and a low-density universe presented a simpler alternative, based on a cosmology containing
matter alone. However, there was another observation that pointed clearly toward the need
for $\Omega_{\Lambda}$: Type Ia supernovae (SNIa) measurements. A detailed discussion of these
measurements is beyond the scope of these lectures, but the principle is simple: SNeIa represent
a {\em standard candle}, i.e. objects whose intrinsic brightness we know, based on observations
of nearby supernovae. They are also extremely bright, so they can be observed at cosmological
distances. Two collaborations, the Supernova Cosmology Project \cite{SCPSNIa} and the High-z 
Supernova Search \cite{HZSNIa} obtained samples of supernovae at redshifts around $z = 0.5$.
This is far enough out that it is possible to measure deviations from the linear Hubble law
$v = H_0 d$ due to the time-evolution of the Hubble parameter: the groups were able to 
{\em measure} the acceleration or deceleration of the universe directly. If the universe is 
decelerating, objects at a given redshift will be closer to us,
and therefore brighter than we would expect based on a linear Hubble law. Conversely, if
the expansion is accelerating, objects at a given redshift will be further away, and therefore
dimmer. The result from both groups was that the supernovae were consistently dimmer than
expected. Fig. \ref{figSCP} shows the data from the Supernova Cosmology Project \cite{SCPHubblediag},
who quoted
a best fit of $\Omega_{\rm M} \simeq 0.3$, $\Omega_{\Lambda} \simeq 0.7$, just what was
needed to reconcile the dynamical mass measurements with a flat universe!

So we have arrived at a very curious picture indeed of the universe: matter, including
both baryons and the mysterious dark matter (which I will not discuss in any detail in 
these lectures)
makes up only about 30\% of the energy density in the universe. The remaining 70\% is
made of of something that looks very much like Einstein's ``greatest blunder'', a 
cosmological constant. This {\em dark energy} can possibly be identified with the vacuum
energy predicted by quantum field theory, except that the energy density is 120 orders
of magnitude smaller than one would expect from a naive analysis. Few, if any, satisfying
explanations have been proposed to resolve this discrepancy. For example, some authors
have proposed arguments based on the Anthropic Principle \cite{anthrop} to explain the low
value of $\rho_{\Lambda}$, but this explanation is controversial to say the least. There
is a large body of literature devoted to the idea that the dark energy is something other
than the quantum zero-point energy we have considered here, the most popular of
which are  self-tuning scalar field models dubbed {\em quintessence} \cite{quintessence}. A
review can be found in Ref. \cite{darkenergyreview}. However, it
is safe to say that the dark energy that dominates the universe is currently unexplained,
but it is of tremendous interest from the standpoint of fundamental theory. This will
form the main theme of these lectures: cosmology provides us a way to study a question
of central importance for particle theory, namely the nature of the vacuum in quantum
field theory. This is something that cannot be studied in particle accelerators, so 
in this sense cosmology provides a unique window on particle physics. We will see later,
with the introduction of the idea of inflation, that vacuum energy is important not
only in the universe today. It had an important influence on the very early universe
as well, providing an explanation for the origin of the primordial density fluctuations
that later collapsed to form all structure in the universe. This provides us with yet
another way to study the ``physics of nothing'', arguably one of the most important
questions in fundamental theory today.

\section{The Cosmic Microwave Background}
\label{seccmb}

In this section we will discuss the background of relic photons in the universe, or 
cosmic microwave background, discovered by Penzias and Wilson at Bell Labs in 1963.
The discovery of the CMB was revolutionary, providing concrete evidence for the Big Bang model
of cosmology over the Steady State model. More precise measurements of the CMB are providing
a wealth of detailed information about the fundamental parameters of the universe.

\subsection{Recombination and the formation of the CMB}

The basic picture of an expanding, cooling universe leads to a number of 
startling predictions: the formation of nuclei and the resulting primordial 
abundances of elements, and the later formation of neutral atoms and the 
consequent presence of a cosmic background of photons, the cosmic microwave 
background (CMB) \cite{CMBreview,CMBbib}. A rough history of the universe can be 
given as a time line of
increasing time and decreasing temperature \cite{kolbturner}:
\begin{itemize}
\item{$T \sim 10^{15}\ K$, $t \sim 10^{-12}\ {\rm sec}$: Primordial soup of
fundamental particles.}
\item{$T \sim 10^{13}\ K$, $t \sim 10^{-6}\ {\rm sec}$: Protons and neutrons form.}
\item{$T \sim 10^{10}\ K$, $t \sim 3\ {\rm min}$: Nucleosynthesis: nuclei form.}
\item{$T \sim 3000\ K$, $t \sim 300,000\ {\rm years}$: Atoms form.}
\item{$T \sim 10\ K$, $t \sim 10^{9}\ {\rm years}$: Galaxies form.}
\item{$T \sim 3\ K$, $t \sim 10^{10}\ {\rm years}$: Today.}
\end{itemize}
The epoch at which atoms form, when the universe was at an age of 300,000 years 
and a temperature of around $3000\ {\rm K}$  is somewhat oxymoronically referred to as 
``recombination'', despite the fact that electrons and nuclei had never before 
``combined'' into atoms. The physics is simple: at a temperature of greater 
than about $3000\ {\rm K}$, the universe consisted of an ionized plasma of mostly 
protons, electrons, and photons, which a few helium nuclei and a tiny trace of 
Lithium. The important characteristic of this plasma is that it was {\it 
opaque}, or more precisely the mean free path of a photon was a great deal 
smaller than the horizon size of the universe. As the universe cooled and 
expanded, the plasma ``recombined'' into neutral atoms, first the helium, then a 
little later the hydrogen. 
\begin{figure}[ht]
\vskip0.5cm
\centerline{\epsfig{file=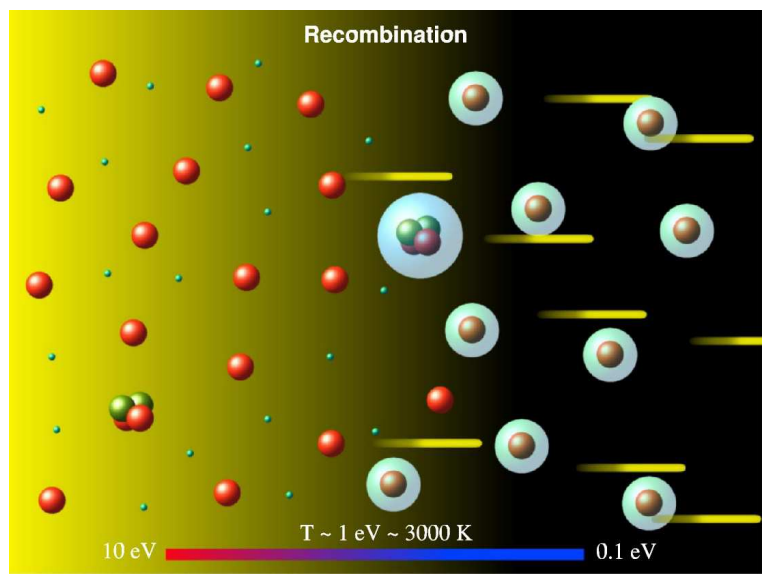, width=10.0cm}}
\caption{Schematic diagram of recombination.}
\label{recomb}
\end{figure}

If we consider hydrogen alone, the process of recombination can be           
described by the {Saha equation} for the  equilibrium ionization fraction 
$X_{\rm e}$ of the hydrogen \cite{sahaeq}:
\begin{equation}
{1 - X_{\rm e} \over X_{\rm e}^2} = {4 \sqrt{2} \zeta(3) \over \sqrt{\pi}} \eta 
\left({T \over m_{\rm e}}\right)^{3/2} \exp\left({13.6\ {\rm eV} \over 
T}\right).\label{eqsahaequation}
\end{equation}
Here $m_{\rm e}$ is the electron mass and $13.6\ {\rm eV}$ is the ionization 
energy of hydrogen. The physically important parameter affecting recombination 
is the density of protons and electrons compared to photons. This is determined
by the {\it baryon  asymmetry}, or the excess of baryons over antibaryons in the 
universe.\footnote{If there were no excess of baryons over antibaryons, there would be no 
protons and
electrons to recombine, and the universe would be just a gas of photons and neutrinos!} 
This is described as the ratio of baryons to photons:
\begin{equation}
\eta \equiv {n_{\rm b} - n_{\rm \bar b} \over n_\gamma} = 2.68 \times 10^{-8} 
\left(\Omega_{\rm b} h^2\right).
\end{equation}
Here $\Omega_{\rm b}$ is the baryon density and $h$ is the Hubble constant in units
of $100\ {\rm km/s/Mpc}$,
\begin{equation}
h \equiv H_0 / (100\ {\rm km /s/Mpc}).
\end{equation}
Recombination happens quickly (i.e., in much less than a Hubble time $t \sim H^{-1}$), but is 
not instantaneous. The universe goes from a completely ionized state to
a neutral state over a range of redshifts $\Delta z \sim 200$. 
If we define recombination as an ionization fraction $X_{\rm e} = 0.1$, we have
that the temperature at recombination $T_{\rm R} = 0.3\ {\rm eV}$.

What happens to the photons after recombination? Once the gas in the universe
is in a neutral state, the mean free path for a photon rises to much larger than
the Hubble distance. The universe is then full of a background of freely propagating
photons with a blackbody distribution of frequencies. At the time of recombination,
the background radiation has a temperature of $T = T_{\rm R} = 3000\ {\rm K}$, and as the
universe expands the photons redshift, so that the temperature of the photons drops
with the increase of the scale factor, $T \propto a(t)^{-1}$. We can detect these
photons today. Looking at the sky, this background of photons comes to us evenly from
all directions, with an observed temperature of $T_0 \simeq 2.73\ {\rm K}$. 
This allows us to determine the redshift of the last scattering surface,
\begin{equation}
1 + z_{\rm R} = {a\left(t_0\right) \over a\left(t_{\rm R}\right)} = {T_{\rm R} \over T_0} \simeq 1100.
\end{equation}
This is the cosmic microwave
background. Since by looking at higher and higher redshift objects, we are looking
further and further back in time, we can view the observation of CMB photons as imaging
a uniform ``surface of last scattering'' at a redshift of 1100.
\begin{figure}[ht]
\vskip0.5cm
\centerline{\epsfig{file=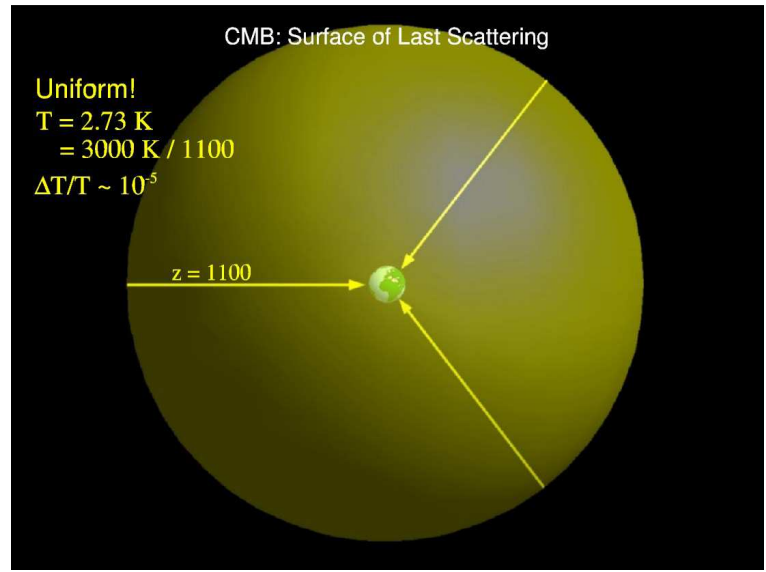, width=10.0cm}}
\caption{Cartoon of the last scattering surface. From earth, we see microwaves radiated
uniformly from all directions, forming a ``sphere'' at 
redshift $z = 1100$.}
\label{figlss}
\end{figure}
To the extent that recombination happens at the same time and in the same way everywhere,
the CMB will be of precisely uniform temperature. In fact the CMB is observed to be
of uniform temperature to about 1 part in 10,000! We shall consider the puzzles presented
by this curious isotropy of the CMB later. 

\begin{figure}[ht]
\vskip0.5cm
\centerline{\epsfig{file=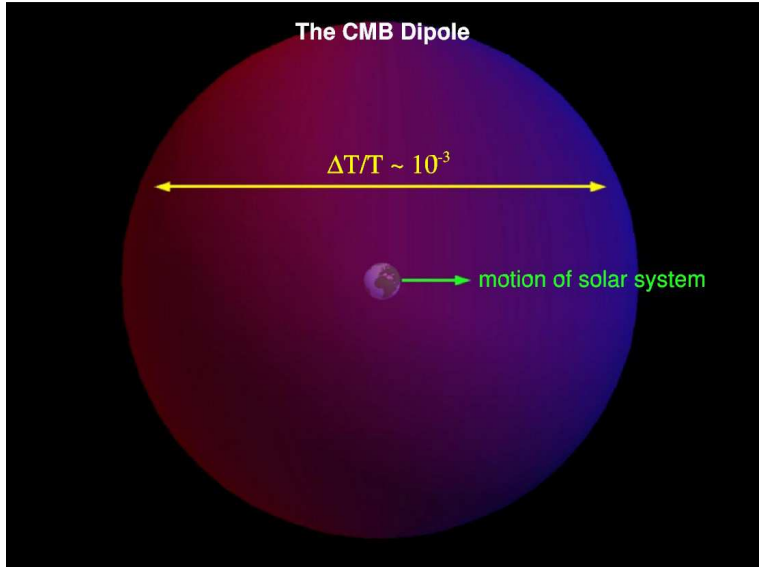, width=10.0cm}}
\caption{The CMB dipole due to the earth's peculiar motion.}
\label{figCMBdipole}
\end{figure}

While the observed CMB is highly isotropic, it is not perfectly so. The largest 
contribution to the anisotropy of the CMB as seen from earth is simply Doppler shift
due to the earth's motion through space. (Put more technically, the motion is the
earth's motion relative to a ``comoving'' cosmological reference frame.) CMB photons 
are slightly blueshifted in the direction of our motion and slightly redshifted 
opposite the direction of our motion. This blueshift/redshift shifts the temperature
of the CMB so the effect has the characteristic form of a ``dipole'' temperature
anisotropy, shown in Fig. \ref{figCMBdipole}.
This dipole anisotropy was first observed in the 1970's by
David T. Wilkinson and Brian E. Corey at Princeton, and another group consisting
of George F. Smoot, Marc V. Gorenstein and Richard A. Muller. They found a dipole
variation in the CMB temperature of about $0.003\ {\rm K}$, or
$(\Delta T / T) \simeq 10^{-3}$, corresponding to a peculiar velocity of the 
earth of about $600\ {\rm km / s}$, roughly in the direction of the constellation Leo.

\begin{figure}[ht]
\vskip0.5cm
\centerline{\epsfig{file=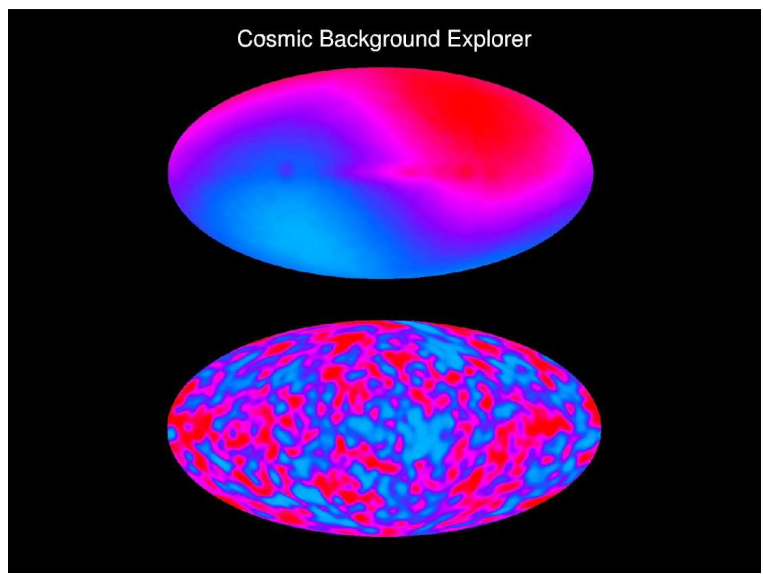, width=10.0cm}}
\caption{The COBE measurement of the CMB anisotropy \cite{refCOBE}. The top oval is a 
map of the sky 
showing the dipole anisotropy $\Delta T / T \sim 10^{-3}$. The bottom oval is a similar
map with the dipole contribution subtracted, showing the anisotropy for $\ell > 1$,
$\Delta T / T \sim 10^{-5}$. (Figure courtesy of the COBE Science Working Group.)}
\label{figCOBE}
\end{figure}

The dipole anisotropy, however, is a {\em local} phenomenon. Any intrinsic, or primordial,
anisotropy of the CMB is potentially of much greater cosmological interest. To describe 
the anisotropy of the CMB, we remember that the surface of last scattering appears to us
as a spherical surface at a redshift of $1100$. Therefore the natural parameters to
use to describe the anisotropy of the CMB sky is as an expansion in spherical harmonics 
$Y_{\ell m}$:
\begin{equation}
{\Delta T \over T} = \sum_{\ell = 1}^{\infty} \sum_{m = -\ell}^{\ell}{a_{\ell m} 
Y_{\ell m}\left(\theta,\phi\right)}.
\end{equation}
Since there is no preferred direction in the universe, the physics is independent
of the index $m$, and we can define
\begin{equation}
C_{\ell} \equiv \sum_{m}{\left\vert a_{\ell m} \right\vert^2}.
\end{equation}
The $\ell = 1$ contribution is just the dipole anisotropy, 
\begin{equation}
\left({\Delta T \over T}\right)_{\ell = 1} \sim 10^{-3}.
\end{equation}
It was not until more than a decade after the discovery of the dipole anisotropy
that the first observation was made of anisotropy for $\ell \geq 2$, by the 
differential microwave radiometer aboard the Cosmic Background Explorer (COBE) 
satellite \cite{refCOBE}, launched in in 1990. COBE observed that the anisotropy at the quadrupole
and higher $\ell$ was two orders of magnitude smaller than the dipole:
\begin{equation}
\left({\Delta T \over T}\right)_{\ell > 1} \simeq 10^{-5}.
\end{equation}
Fig. \ref{figCOBE} shows the dipole and higher-order CMB anisotropy as measured by
COBE. 
It is believed that this anisotropy represents intrinsic fluctuations in the CMB
itself, due to the presence of tiny primordial density fluctuations in the cosmological
matter present at the time of recombination. These density fluctuations are of great
physical interest, first because these are the fluctuations which later collapsed
to form all of the structure in the universe, from superclusters to planets to
graduate students. Second, we shall see that within the paradigm of inflation, the
form of the primordial density fluctuations forms a powerful probe of the physics
of the very early universe. The remainder of this section will be concerned with
how primordial density fluctuations create fluctuations in the temperature of the CMB. 
Later on, I will discuss using the CMB as a tool to probe other physics, especially
the physics of inflation.

While the physics of recombination in the homogeneous case is quite simple, the
presence of inhomogeneities in the universe makes the situation much more complicated. 
I will describe some of the major effects qualitatively here, and refer the reader
to the literature for a more detailed technical explanation of the relevant
physics \cite{CMBreview}. The complete linear theory of CMB fluctuations was
first worked out by Bertschinger and Ma in 1995 \cite{ma95}. 

\subsection{Sachs-Wolfe Effect}
The simplest contribution to the CMB anisotropy from density fluctuations is just
a gravitational redshift, known as the {\em Sachs-Wolfe effect} \cite{sachswolfe}. 
A photon coming from a region which is slightly overdense will have a slightly larger
redshift due to the deeper gravitational well at the surface of last scattering. 
Conversely, a photon coming from an underdense region will have a slightly smaller
redshift. Thus we can calculate the CMB temperature anisotropy due to the slightly
varying Newtonian potential $\Phi$ from density fluctuations at the surface
of last scattering:
\begin{equation}
\left({\Delta T \over T}\right) = {1 \over 3} \Phi,
\end{equation}
where the factor $1/3$ is a general relativistic correction. Fluctuations on large
angular scales (low multipoles) are actually larger than the horizon at the time
of last scattering, so that this essentially kinematic contribution to the CMB 
anisotropy is dominant on large angular scales. 

\subsection{Acoustic oscillations and the horizon at last scattering}

\begin{figure}[ht]
\vskip0.5cm
\centerline{\epsfig{file=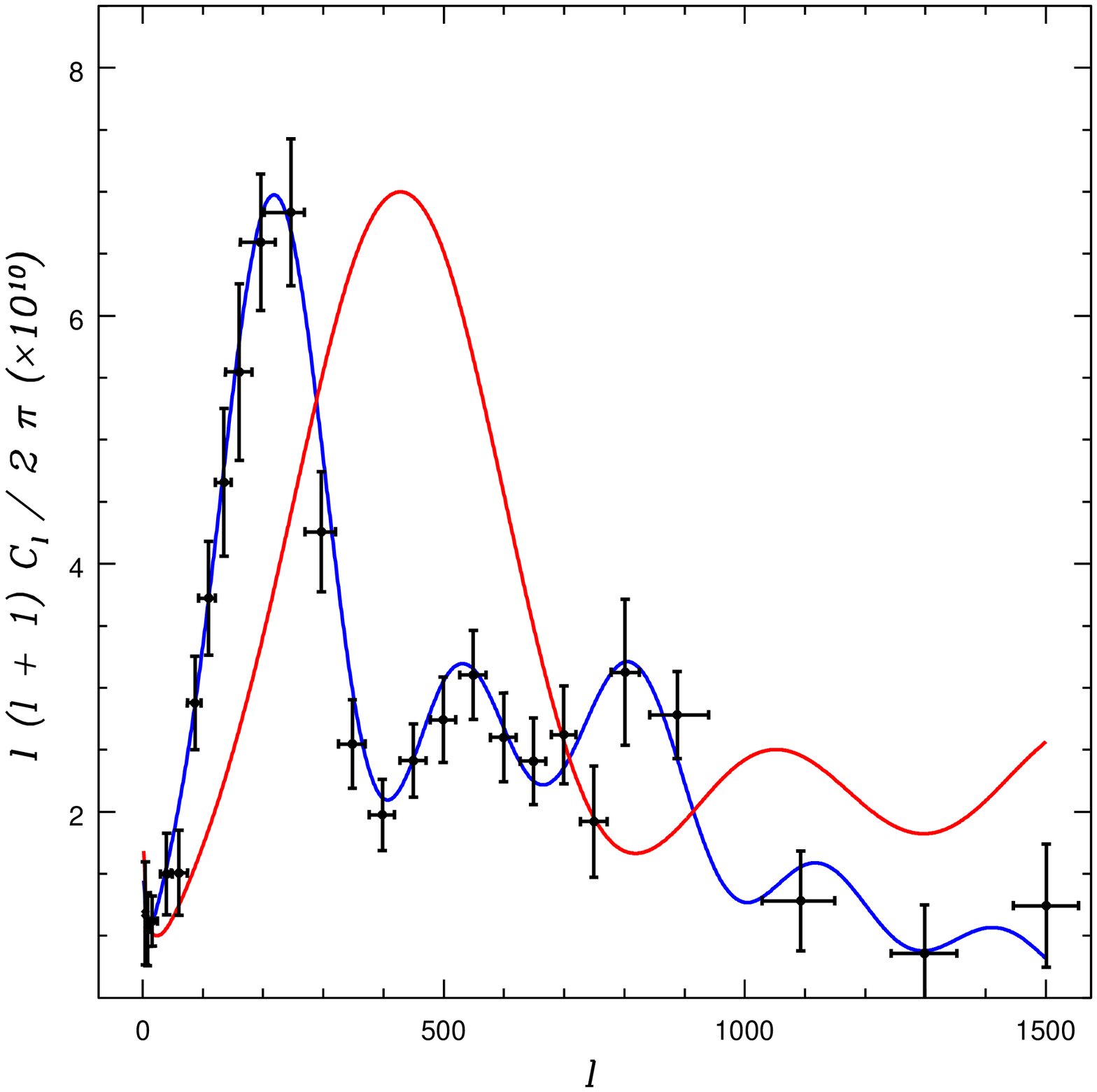, width=10.0cm}}
\caption{$C_{\ell}$ spectra for a universe with $\Omega_{\rm M} = 0.3$ and $\Omega_{\Lambda} = 0.7$
(blue line) and for $\Omega_{\rm M} = 0.3$ and $\Omega_{\Lambda} = 0$ (red line). The open
universe is conclusively ruled out by the current data \cite{wang02} (black crosses).}
\label{figClflatopen}
\end{figure}

For fluctuation modes smaller than the horizon size, more complicated physics comes into
play. Even a summary of the many effects that determine the precise shape of the CMB 
multipole spectrum is beyond the scope of these lectures, and the student is referred
to Refs. \cite{CMBreview} for a more detailed discussion. However, the dominant process
that occurs on short wavelengths is important to us. These are {\em acoustic oscillations}
in the baryon/photon plasma. The idea is simple: matter tends to collapse due to gravity 
onto regions where the density is higher than average, so the baryons ``fall'' into overdense
regions. However, since the baryons and the photons are still strongly coupled, the photons
tend to resist this collapse and push the baryons outward. The result is ``ringing'', or
oscillatory modes of compression and rarefaction in the gas due to density fluctuations. The
gas heats as it compresses and cools as it expands, and this creates fluctuations in the
temperature of the CMB. This manifests itself in the $C_\ell$ spectrum as a series of
bumps (Fig. \ref{figClflatopen}). The specific shape and location of the bumps is created
by complicated, although well-understood physics, involving a large number of cosmological
parameters. The shape of the CMB multipole spectrum depends, for example, on the baryon
density $\Omega_{\rm b}$, the Hubble constant $H_0$, the densities of matter $\Omega_{\rm M}$ and
cosmological constant $\Omega_{\Lambda}$, and the
amplitude of primordial gravitational waves (see Sec. \ref{secinflationflucts}). This
makes interpretation of the spectrum something of a complex undertaking, but it also
makes it a sensitive probe of cosmological models. In these lectures, I will primarily
focus on the CMB as a probe of inflation, but there is much more to the story.

\begin{figure}[ht]
\vskip0.5cm
\centerline{\epsfig{file=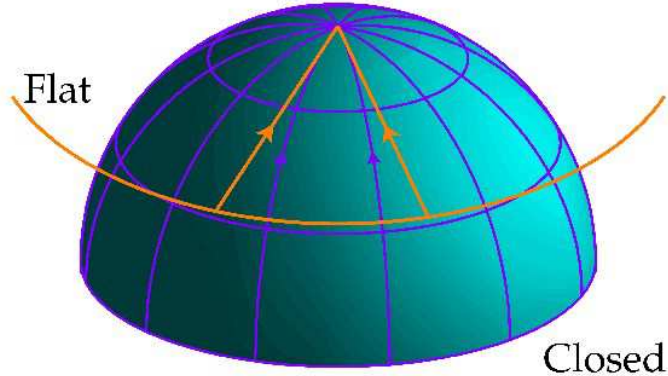, width=10.0cm}}
\caption{The effect of geometry on angular size. Objects of a given angular size are smaller
in a closed space than in a flat space. Conversely, objects of a given angular size are
larger in an open space. (Figure courtesy of Wayne Hu \cite{waynehuwebsite}.)}
\label{figgeometry}
\end{figure}

\begin{figure}[ht]
\vskip0.5cm
\centerline{\epsfig{file=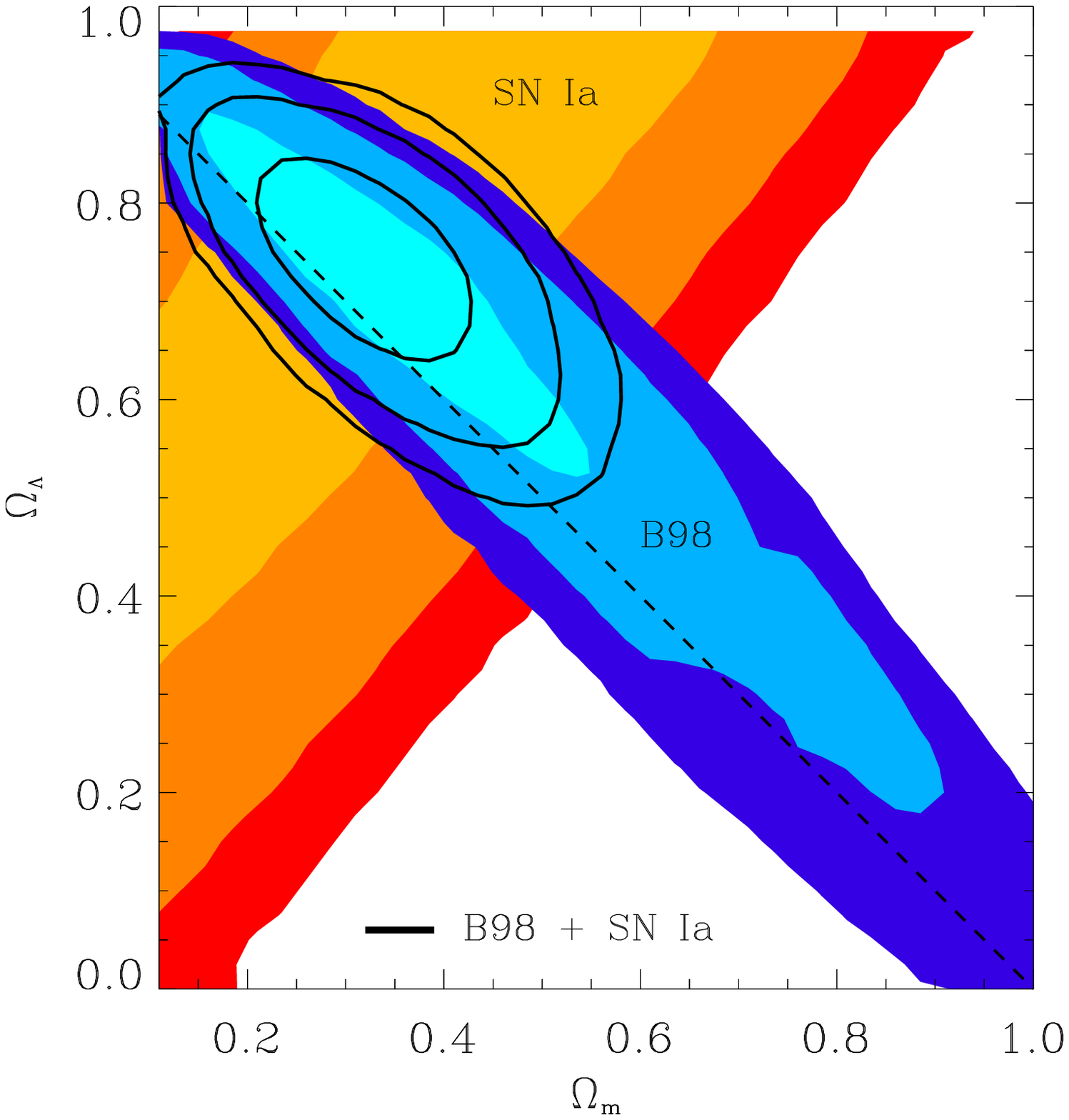, width=10.0cm}}
\caption{Limits on $\Omega_{\rm M}$ and $\Omega_{\rm \Lambda}$ from the CMB and from
Type Ia supernovae. The two data sets together strongly favor a flat universe (CMB)
with a cosmological constant (SNIa). \cite{CMBSNIacombined}}
\label{figOmOL}
\end{figure}

These oscillations are sound waves in the direct sense: compression waves in the gas.
The position of the bumps in $\ell$ is determined by the oscillation frequency of the mode.
The first bump is created by modes that have had time to go through half an oscillation in
the age of the universe (compression), the second bump modes that have gone through
one full oscillation (rarefaction), and so on. So what is the wavelength of a mode that
goes through half an oscillation in a Hubble time? About the horizon size at the time
of recombination, 300,000 light years or so! This is an immensely powerful tool: it 
in essence provides us with a ruler of known length (the wavelength of the oscillation
mode, or the horizon size at recombination), situated at a known distance (the distance
to the surface of last scattering at $z = 1100$). The angular size of this ruler when
viewed at a fixed distance depends on the curvature of the space that lies between us
and the surface of last scattering (Fig. \ref{figgeometry}). 
If the space is  negatively curved, the ruler will
subtend a smaller angle than if the space is flat;\footnote{To paraphrase Gertrude Stein,
``there's more {\em there} there.''} if the space is positively curved, the
ruler will subtend a larger angle. We can measure the ``angular size'' of our ``ruler''
by looking at where the first acoustic peak shows up on the plot of the $C_{\ell}$ spectrum
of CMB fluctuations. The positions of the peaks are determined by the curvature of the
universe.\footnote{Not surprisingly, the real situation is a good deal more complicated than what
I have described here. \cite{CMBreview}}. This is how we measure $\Omega$ with the
CMB. Fig. \ref{figClflatopen} shows an $\Omega = 1$ model and an $\Omega = 0.3$ model
along with the current data. The data allow us to clearly distinguish between flat and
open universes. Figure \ref{figOmOL} shows limits from Type Ia supernovae and the CMB
in the space of $\Omega_{\rm M}$ and $\Omega_{\Lambda}$. 

\section{Inflation}
\label{secinflation}

The basic picture of Big Bang cosmology, a hot, uniform early universe expanding
and cooling at late times, is very successful and has (so far) passed a battery of
increasingly precise tests. It successfully explains the observed primordial 
abundances of elements, the observed redshifts of distant galaxies, and the presence
of the cosmic microwave background. Observation of the CMB is a field that is currently
progressing rapidly, allowing for extremely precise tests of cosmological models. The 
most striking feature of the CMB is its high degree of uniformity, with inhomogeneities
of about one part in $10^5$. Recent precision measurements of the tiny anisotropies in
the CMB have allowed for constraints on a variety of cosmological parameters. Perhaps
most importantly, observations of the first acoustic peak, first accomplished with precision 
by the Boomerang \cite{boomerang} and MAXIMA \cite{maxima} experiments, indicate that the 
geometry of the universe is flat, with 
$\Omega_{\rm total} = 1.02 \pm 0.05$ \cite{CMBSNIacombined}. However, this success of the 
standard Big
Bang leaves us with a number of vexing puzzles. In particular, how did the universe
get so big, so flat, and so uniform? We shall see that these observed characteristics
of the universe are poorly explained by the standard Big Bang scenario, and we will
need to add something to the standard picture to make it all make sense: inflation.

\subsection{The flatness problem}
\label{secflatnessproblem}

We observe that the universe has a nearly flat geometry, $\Omega_{\rm tot} \simeq 1$. 
However, this is far from a natural expectation for an arbitrary FRW
space. It is simple to see why. Take the defining expression for $\Omega$,
\begin{equation}
\Omega \equiv {8 \pi G \over 3} \left({\rho \over H^2}\right).
\end{equation}
Here the density of matter with equation of state $p = w \rho$ evolves with expansion
as
\begin{equation}
\rho \propto a^{-3 (1 + w)}.
\end{equation}
Using this and the Friedmann equation (\ref{eqRW}) it is possible to derive a simple 
expression for how $\Omega$ evolves with expansion:
\begin{equation}
\label{eqflatnessproblem}
{d \Omega \over d \log{a}} = (1 + 3 w) \Omega \left(\Omega - 1\right).
\end{equation}
This is most curious! Note the sign. For an equation of state with $1 + 3 w > 0$, 
which is the case for any kind of ``ordinary'' matter, a flat universe is an unstable
equilibrium:
\begin{equation}
{d \left\vert \Omega - 1 \right\vert \over d \log{a}} > 0,\ 1 + 3 w > 0.
\end{equation}
So if the universe at early times deviates even slightly from a flat geometry, that
deviation will grow large at late times.  If the universe today is flat to within
$\Omega \simeq 1 \pm 0.05$, then $\Omega$ the time of recombination was $\Omega = 1 \pm 0.00004$,
and at nucleosynthesis $\Omega = 1 \pm 10^{-12}$. This leaves unexplained how the universe
got so  perfectly flat in the first place. This curious fine-tuning in cosmology is referred
to as the {\em flatness problem}. 

\subsection{The horizon problem}
\label{sechorizonproblem}

\begin{figure}[ht]
\vskip0.5cm
\centerline{\epsfig{file=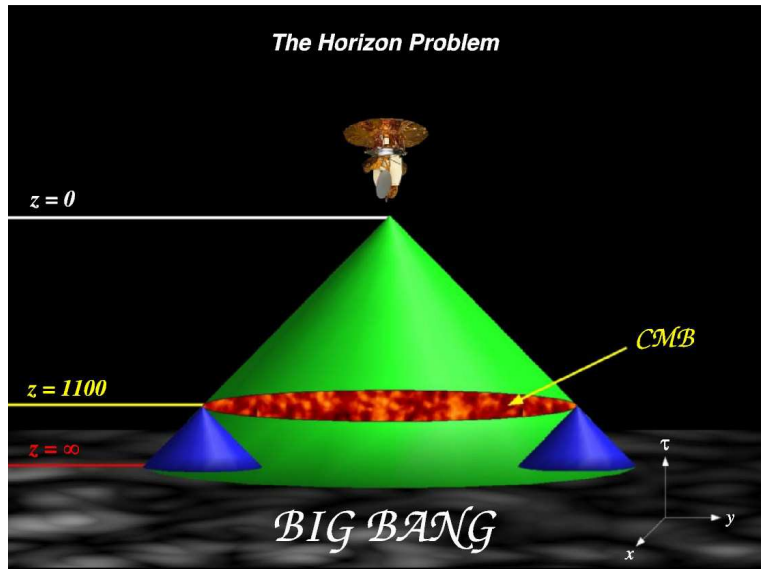, width=10.0cm}}
\caption{Light cones in a FRW space, plotted in conformal coordinates, 
$ds^2 = a^2 \left(d\tau^2 - d{\bf x}^2\right)$. The ``Big Bang'' is a surface at $\tau = 0$. 
Points in the space can only be in causal contact if their past light cones intersect at the
Big Bang.}
\label{fighorizonproblem}
\end{figure}

There is another odd fact about the observed universe: the apparent high degree of uniformity
and thermal equilibrium of the CMB. While it might at first seem quite natural for the
hot gas of the early universe to be in good thermal equilibrium, this is in fact quite 
{\em unnatural} in the standard Big Bang picture. This is because of the presence of a
cosmological horizon. We recall that the horizon size of the universe is just how far a
photon can have traveled since the Big Bang, $d_{\rm H} \sim t$ in units where $c = 1$. This
defines how large a ``patch'' of the universe can be in causal contact. Two points in the
universe separated by more than a horizon size have no way to reach thermal equilibrium,
since they cannot have {\em ever} been in causal contact. Consider two points 
comoving with respect to the cosmological expansion. The physical distance $d$ 
between the two points then just increases linearly with the scale factor:
\begin{equation}
d \propto a\left(t\right).
\end{equation}
The horizon size is just proportional to the time, or equivalently to the inverse of the
Hubble parameter,
\begin{equation}
d_{\rm H} \propto H^{-1}.
\end{equation}
It is straightforward to show that for any FRW space with constant equation of state, 
there is a conserved quantity given by
\begin{equation}
\label{eqFRWconserved}
\left({d \over d_{\rm H}}\right)^2 \left\vert \Omega - 1 \right\vert = {\rm const.}
\end{equation}
Proof of this is also left as an exercise for the student. 
If we again consider the case of ``ordinary'' matter, with equation of state $1 + 3 w > 0$,
we see from Eqs. (\ref{eqflatnessproblem}) and (\ref{eqFRWconserved}) that the horizon 
expands faster than a comoving length:
\begin{equation}
{d \over d \log{a}}\left({d \over d_{\rm H}}\right) < 0,\ 1 + 3 w > 0.
\end{equation}
This means that any two points at rest relative to the expansion of the universe will
be causally disconnected at early times and causally connected at late 
times. This can be visualized on a conformal space-time diagram 
(Fig. \ref{fighorizonproblem}). On a conformal diagram, we re-write the metric in terms of
{\em conformal time} $d\tau \equiv dt / a$:
\begin{equation}
ds^2 = a^2\left(\tau\right) \left[d\tau^2 - d {\bf x}^2\right].
\end{equation}
The advantage of this choice of coordinates is that light always travels at 
$45^\circ$ angles $d {\bf x} = d\tau$ on the diagram, regardless 
of the behavior of the scale factor. In a matter dominated universe, the scale factor
evolves as
\begin{equation}
a\left(\tau\right) \propto \tau^2,
\end{equation}
so the Big Bang, $a = 0$, is a surface ${\bf x} = {\rm const.}$ at $\tau = 0$. Two points on a 
given $\tau = {\rm const.}$ surface are in causal contact if their past light cones intersect at 
the Big Bang, $\tau = 0$.

So the natural expectation for the very early universe is that there should be
a large number of small, causally disconnected regions that will be in poor thermal
equilibrium. 
The central question is: just how large was the horizon when the CMB
was emitted? Since the universe was about 300,000 years old at recombination, the
horizon size then was about 300,000 light years. Each atom at the surface of last 
scattering could only be in causal contact (and therefore in thermal equilibrium) with
with other atoms within a radius of about 300,000 light years. As seen from earth,
the horizon size at the surface of last scattering subtends an angle on the sky of
about a degree.
\begin{figure}[ht]
\vskip0.5cm
\centerline{\epsfig{file=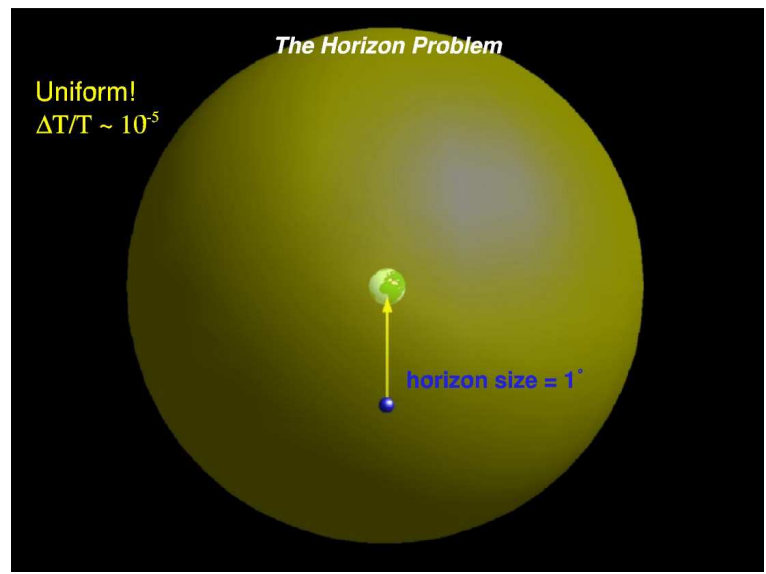, width=10.0cm}}
\caption{Schematic diagram of the horizon size at the surface of last scattering. The
horizon size at the time of recombination was about 300,000 light years. Viewed from
earth, this subtends an angle of about a degree on the sky.}
\label{fighorizonlss}
\end{figure}
Therefore, two points on the surface of last scattering separated by an angle of more
than a degree were out of causal contact at the time the CMB was emitted. However, the
CMB is uniform (and therefore in thermal equilibrium) over the {\em entire sky} to 
one part in $10^{5}$. How did all of these disconnected regions reach such a high degree
of thermal equilibrium?

\subsection{Inflation}

The flatness and horizon problems have no solutions within the context of a standard
matter- or radiation-dominated universe, since for any ordinary matter, the force
of gravity causes the expansion of the universe to decelerate. The only available fix 
would appear to be
to invoke initial conditions: the universe simply started out flat, hot, and in thermal
equilibrium. While this is certainly a possibility, it hardly a satisfying explanation.
It would be preferable to have an explanation for {\em why} the universe was flat,
hot, and in thermal equilibrium. Such an explanation was proposed by Alan Guth in 
1980 \cite{guth80} under the name of {\em inflation}. Inflation is the idea that at
some very early epoch, the expansion of the universe was accelerating instead of
decelerating.\footnote{Similar ideas had been discussed in the literature for some time \cite{earlyinflationpapers}.}

Accelerating expansion turns the horizon and flatness problems on their heads. This
is evident from the equation for the acceleration,
\begin{equation}
{\ddot a \over a} = - (1 + 3 w) \left({4 \pi G\over 3} \rho\right).
\end{equation}
We see immediately that the condition for acceleration $\ddot a > 0$ is that the
equation of state be characterized by negative pressure, $1 + 3 w < 0$. This 
means that the universe evolves {\em toward} flatness rather than away:
\begin{equation}
{d \left\vert \Omega - 1 \right\vert \over d \log{a}} < 0,\ 1 + 3 w < 0.
\end{equation}
Similarly, from Eq. (\ref{eqFRWconserved}), we see that comoving scales grow in size
more quickly than the horizon,
\begin{equation}
{d \over d \log{a}}\left({d \over d_{\rm H}}\right) > 0,\ 1 + 3 w < 0.
\end{equation}
This is a very remarkable behavior. It means that two points that are initially in causal
contact ($d < d_{\rm H}$) will expand so rapidly that they will eventually be
causally {\em disconnected}. Put another way, two points in space whose relative 
velocity due to expansion is less than the speed of light will eventually be flying
apart from each other at greater than the speed of light! Note that there is absolutely
no violation of the principles of relativity. Relative velocities $v > c$ are allowed
in general relativity as long as the observers are sufficiently separated in space.\footnote{
An interesting consequence of the currently observed accelerating expansion 
is that all galaxies except those in our local group will eventually be moving away from us
faster than the speed of light and will be invisible to us. The far future universe
will be a lonely place, and cosmology will be all but impossible!}
This mechanism provides a neat way to explain the apparent homogeneity of the universe
on scales much larger than the horizon size: a tiny region of the universe, initially
in some sort of equilibrium, is ``blown up'' by accelerated expansion to an enormous
and causally disconnected scale. 

\begin{figure}[ht]
\vskip0.5cm
\centerline{\epsfig{file=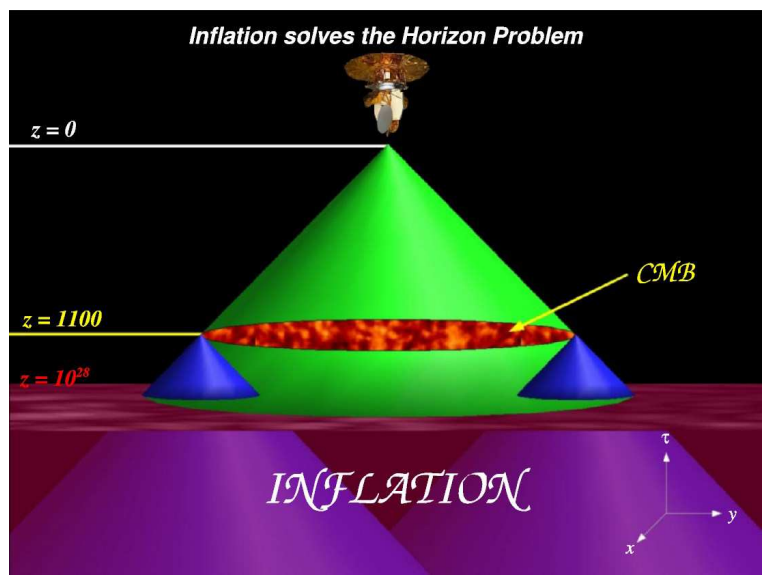, width=10.0cm}}
\caption{Conformal diagram of light cones in an inflationary space. The end of inflation 
creates an ``apparent'' 
Big Bang at  $\tau = 0$ which is at high (but not infinite) redshift. There is, however,
no singularity at $\tau = 0$ and the light cones intersect at an earlier time.}
\label{figinflationhorizon}
\end{figure}

We can plot the inflationary universe on a conformal 
diagram (Fig. \ref{figinflationhorizon}). 
In an inflationary universe, there is no singularity $a \rightarrow 0$ at conformal time
$\tau = 0$. To see this, take the case of a cosmological constant, $\rho_\Lambda = {\rm const.}$,
which corresponds to exponential increase of the scale factor:
\begin{equation}
a\left(t\right) \propto e^{H t},\ H = {\rm const.}
\end{equation}
In this case, the conformal time is {\em negative} and evolves toward zero with increasing
``proper'' time $t$:
\begin{equation}
\tau = \int{dt \over a (t)} \propto -{1 \over H} e^{-H t},
\end{equation}
so that the scale factor evolves as 
\begin{equation}
\label{eqinflscalefactor}
a\left(\tau\right) = - {1 \over H \tau}.
\end{equation}
The scale factor becomes infinite at $\tau = 0$! This is because we have assumed $H = {\rm const.}$,
which means that inflation will continue forever, with $\tau = 0$ representing the infinite future,
$t \rightarrow \infty$. In the real universe, inflation ends at some finite time, and the
approximation (\ref{eqinflscalefactor}), while valid at early times, breaks down near the end
of inflation. 
So the surface $\tau = 0$ is not the Big Bang, but the end of inflation. The initial singularity
has been pushed back arbitrarily far in conformal time $\tau \ll 0$, and light cones can 
extend through the apparent ``Big Bang'' so that apparently disconnected points are in 
causal contact.

How much inflation do we need to solve the horizon and flatness problems? We will see that 
sensible models of inflation tend to place the inflationary epoch at a time when the temperature
of the universe was typical of Grand Unification,
\begin{equation}
T_{\rm i} \sim  10^{15}\ {\rm GeV},
\end{equation}
so that the horizon size, or size of a causal region, was about
\begin{equation}
d_{\rm H}\left(t_{\rm i}\right) \sim H^{-1} \sim {m_{\rm Pl} \over T_{\rm i}^2} \sim 10^{-11}\ {\rm GeV}^{-1}.
\end{equation}
In order for inflation to solve the horizon problem, this causal region must be blown up to 
{\em at least} the size of the observable universe today,\footnote{Exercise for the student: what 
is 1 km/s/MpC measured in units of GeV?}
\begin{equation}
d_{\rm H}\left(t_0\right) \sim H_0^{-1} \sim 10^{41}\ {\rm GeV}^{-1}.
\end{equation}
So that the scale factor must increase by about
\begin{equation}
{\delta a \over a} \sim \left({a(t_i) \over a(t_0)}\right) {d_{\rm H}\left(t_0\right) \over d_{\rm H}\left(t_{\rm i}\right)} \sim \left({T_0 \over T_i}\right) {d_{\rm H}\left(t_0\right) \over d_{\rm H}\left(t_{\rm i}\right)} \sim 10^{24},
\end{equation}
or somewhere around a factor of $e^{55}$. Here the extra factor $a(t_i) / a(t_0)$ accounts for 
the expansion between the end of inflation $T_{\rm i} \sim 10^{15}\ {\rm GeV}$ and today, 
$T_0 \sim 10^{-4}\ {\rm eV}$. This is the {\em minimum} amount of inflation required
to solve the horizon problem, and inflation can in fact go on for much longer. In the next 
section we will talk about how one constructs a model of inflation in particle theory. A more 
detailed introductory review can be found in Ref. \cite{liddlereview}.

\subsection{Inflation from scalar fields}

We have already seen that a cosmological constant due to nonzero vacuum energy results
in accelerating cosmological expansion. While this is a good candidate for explaining
the observations of Type Ia supernovae, it does not work for explaining inflation at
early times for the simple reason that any period of accelerated expansion in the very
early universe must end. Therefore the vacuum energy driving inflation must be dynamic.
To implement a time-dependent ``cosmological constant,'' we require a field with the
same quantum numbers as vacuum, i.e. a scalar. We will consider a scalar field minimally
coupled to gravity, with potential $V\left(\phi\right)$ and Lagrangian
\begin{equation}
{\cal L} = {1 \over 2} g^{\mu \nu} \partial_\mu \phi \partial_\nu \phi - V\left(\phi\right),
\end{equation}
where we have modified a familiar Minkowski-space field theory by replacing the Minkowski
metric $\eta^{\mu \nu}$ with the FRW metric $g^{\mu \nu}$. The equation of motion for
the field is 
\begin{equation}
\ddot \phi + 3 H \dot\phi + p^2 \phi + V'\left(\phi\right) = 0.
\end{equation}
This is the familiar equation for a free scalar field with an extra piece, $3 H \dot \phi$,
that comes from the use of the FRW metric in the Lagrangian. We will be interested in the
ground state of the field $p = 0$. This is of interest because the zero mode of the
field forms a perfect fluid, with energy density
\begin{equation}
\rho = {1 \over 2} \dot\phi^2 + V\left(\phi\right)
\end{equation}
and pressure
\begin{equation}
p = {1 \over 2} \dot\phi^2 - V\left(\phi\right).
\end{equation}
Note in particular that in the limit $\dot\phi \rightarrow 0$ we recover a cosmological 
constant, $p = -\rho$, as long as the potential $V\left(\phi\right)$ is nonzero. The
Friedmann equation for the dynamics of the cosmology is
\begin{equation}
H^2 = \left({\dot a \over a}\right)^2 = {8 \pi \over 3 m_{\rm Pl}^2} \left[{1 \over 2} \dot\phi^2
+ V\left(\phi\right)\right].
\end{equation}
(Note that we have written Newtons' constant $G$ in terms of the Planck mass, so that
$G = m_{\rm Pl}^{-2}$ in units where $\hbar = c = 1$.) 
In the $\dot\phi \rightarrow 0$ limit, we have 
\begin{equation}
H^2 = {8 \pi \over 3 m_{\rm Pl}^2} V\left(\phi\right) = {\rm const.},
\end{equation}
so that the universe expands exponentially,
\begin{equation}
a(t) \propto e^{H t}.
\end{equation}
This can be generalized to a time-dependent field and a quasi-exponential expansion
in a straightforward way. If we have a slowly varying field $(1/2) \dot\phi^2 \ll V(\phi)$,
we can write the equation of motion of the field as
\begin{equation}
\label{eqslowrollEOM}
3 H \dot\phi + V'\left(\phi\right) \simeq 0,
\end{equation}
and the Friedmann equation as
\begin{equation}
\label{eqslowrollfriedmann}
H^2(t) \simeq {8 \pi \over 3 m_{\rm Pl}^2} V\left[\phi\left(t\right)\right],
\end{equation}
so that the scale factor evolves as
\begin{equation}
a(t) \propto \exp{\int{H dt}}.
\end{equation}
This is known as the {\em slow roll} approximation, and corresponds physically to the
field evolution being dominated by the ``friction'' term $3 H \dot\phi$ in the equation
of motion. This will be the case if the potential is sufficiently flat, $V'(\phi) \ll V(\phi)$.
It is possible to write the equation of state of the field in the slow roll approximation
as 
\begin{equation}
p = \left[{2 \over 3} \epsilon(\phi) - 1\right] \rho,
\end{equation}
where the {\em slow roll parameter} $\epsilon$ is given by
\begin{equation}
\epsilon = {m_{\rm Pl}^2 \over 16 \pi} \left[{V'\left(\phi\right) \over V\left(\phi\right)}\right]^2.
\end{equation}
This parameterization is convenient because the condition for accelerating expansion is
$\epsilon < 1$:
\begin{equation}
{\ddot a \over a} = H^2 \left(1 - \epsilon\right).
\end{equation}
Specifying a model for inflation is then as simple as selecting a potential $V\left(\phi\right)$ 
and evaluating its behavior as a source of cosmological energy density. Many models have been
proposed:  Refs.  \cite{linde, riottoandlyth} contain extensive reviews of
inflationary model building.  We will discuss the observational predictions of various models
in Section \ref{secinflationzoology} below. In the next section, we will discuss one of the
central observational predictions of inflation: the generation of primordial density fluctuations.

\subsection{Density fluctuations from inflation}
\label{secinflationflucts}

So far we have seen that the standard FRW cosmology has two (related) unexplained issues: why
is the universe so flat and why is it so smooth? These questions are difficult to answer in the
context of a matter- or radiation-dominated cosmology without resorting to simply setting the
initial conditions of the universe to match what we see, certainly an unsatisfying solution. 
The idea of {\em inflation}, a period of accelerated expansion in the very early universe,
provides an elegant solution to these problems. At some early time, the energy density of the
universe was dominated by something with an equation of state approximating that of vacuum
energy, $p < - (1/3) \rho$. (In the discussion above, we considered a single scalar field as
a model for such a fluid.) The useful thing about inflation is that it dynamically drives the
universe toward a flat ($\Omega = 1$) geometry, and simultaneously drives causally connected
points to non-causal regions, explaining the homogeneity of the universe on scales larger
than the horizon. This is an appealing scenario, but it is short on observational consequences:
how do we tell whether or not inflation actually happened? 

\begin{figure}[ht]
\vskip0.5cm
\centerline{\epsfig{file=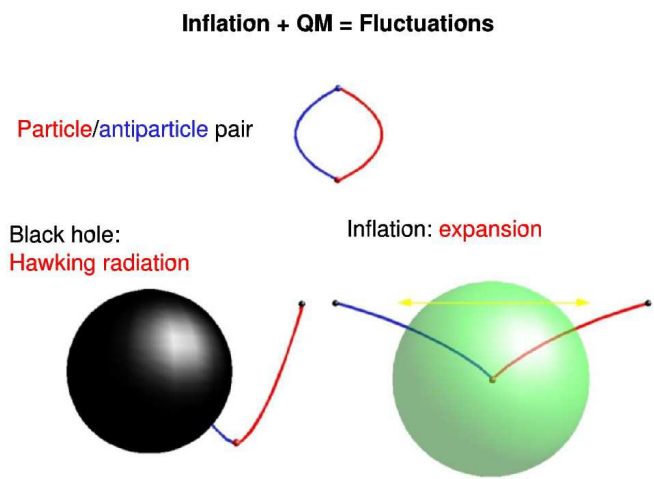, width=10.0cm}}
\caption{Diagrams of virtual particles in a Minkowski space, near the horizon of a black hole,
and in an inflationary space. Like the radiation created by one of the virtual pair being
lost behind the horizon, virtual particles in inflation are swept out of each others' horizon
before they can recombine.}
\label{figvirtualparticles}
\end{figure}

Very soon after the introduction of inflation by Guth, it was realized that inflation had another
remarkable property: it could explain the generation of the primordial density fluctuations
in the universe. This was first worked out independently by Hawking \cite{hawkingquantummodes}, 
Starobinsky \cite{starobquantummodes}, and by Guth \cite{guthquantummodes}. We will cover the 
generation of fluctuations in great detail
later, but the basic physics is familiar, and is closely related to the generation of
Hawking radiation by black holes. It can be explained qualitatively as follows (Fig. 
\ref{figvirtualparticles}). In a normal
Minkowski space, vacuum fluctuations are interpreted as pairs of virtual particles appearing 
and then immediately annihilating as a consequence of the Heisenberg uncertainty principle. 
One qualitative explanation of Hawking radiation at the event horizon of a black hole is
that one of the two virtual particles is trapped by the horizon, leaving the other to escape
as apparently thermal radiation. A similar process holds in an inflationary spacetime: in
inflation, the expansion is so rapid that pairs of virtual particles get ``swept up'' in the
spacetime and are inflated to causally disconnected regions. In essence, they can no longer
find each other to annihilate, and the quantum fluctuations become classical modes of the
field. 
Formally, this effect is calculated by considering the equation of motion for a fluctuation
in a free field $\delta \phi$,
\begin{equation}
\ddot{\left(\delta \phi\right)} + 3 H \dot{\left(\delta \phi\right)} + \left({k \over a}\right)^2 \left(\delta \phi\right) = 0,
\end{equation}
where $k$ is a comoving wavenumber which stays constant with expansion. The physical
momentum $p$ of the particle is
\begin{equation}
{\bf p} = {{\bf k} \over a}.
\end{equation} 
During inflation, the wavelength of a quantum mode is ``stretched'' by the rapid expansion,
\begin{equation}
\lambda \propto a \propto e^{H t}.
\end{equation}
The horizon size, however, remains roughly constant,
\begin{equation}
d_{\rm H} \sim H^{-1} \simeq {\rm const.}.
\end{equation}
Short-wavelength vacuum fluctuations are then quickly redshifted by the expansion until their
wavelengths are larger than the horizon size of the spacetime, and the modes are ``frozen'' as
classical fluctuations. The amplitudes of quantum modes in inflation are conventionally expressed 
at {\em horizon crossing}, that is when the wavelength of the mode is equal to the horizon size,
or $k / a = H$.  The two-point correlation function of the field at horizon crossing is just 
given by the Hubble parameter:
\begin{equation}
\left\langle \delta \phi^2 \right\rangle_{(k/a) = H} = \left({H \over 2 \pi}\right)^2.
\end{equation}
Like the horizon of a black hole, the horizon of an inflationary space has a ``temperature'' 
$T \sim H^{-1}$. The simplest example of the creation of fluctuations during inflation is
the generation of gravitational waves, or {\em tensor modes}. An arbitrary perturbation to
the metric tensor can be expressed as a sum scalar, vector, and tensor fluctuations, depending
on how each behaves under coordinate transformations,
\begin{equation}
\delta g_{\mu \nu} = h^{\rm scalar}_{\mu \nu} + h^{\rm vector}_{\mu \nu} + h^{\rm tensor}_{\mu \nu}.
\end{equation}
The tensor component can be expressed in general as the superposition of two gravitational
wave modes,
\begin{equation}
h^{\rm tensor}_{+,\times} = \varphi_{+,\times} / m_{\rm Pl},
\end{equation}
where $+,\times$ refer to longitudinal and transverse polarization modes, respectively. Each 
mode has the equation of motion of a free scalar during inflation. Therefore fluctuations
in the field are generated by the rapid expansion,
\begin{equation}
\varphi_{+,\times} = {H \over 2 \pi}.
\end{equation}
The case of scalar fluctuations is more complex, since scalar modes are generated by fluctuations
in the inflaton field itself and couple to the curvature of the spacetime. We will simply note
the correct expression,\footnote{The convention for normalizing this expression varies in the
literature.}
\begin{equation}
{\delta \rho \over \rho} = {1 \over 2 \sqrt{\pi}} {\delta N \over \delta \phi} \delta \phi 
                         = {H^2 \over 4 \pi^{3/2} \dot \phi},
\end{equation}
where $N$ is the number of e-folds of inflation,
\begin{equation}
\label{eqnumefoldsdef}
a \propto e^{N} \propto \exp{\int H dt}.
\end{equation}
What about vector modes, or primordial vorticity? Since there is no way to source such modes
with a single scalar field, primordial vector perturbations vanish, at least in simple models
of inflation.

\subsection{The primordial power spectrum}

Summarizing the results of the last section, inflation predicts not only a flat, smooth universe,
but also provides a natural mechanism for the production of primordial density and gravitational
wave fluctuations. The scalar, or density fluctuation amplitude when a mode crosses the horizon
is given by
\begin{equation}
\label{eqscalaramplitude}
\left({\delta \rho \over \rho}\right)_{k = a H} = {H^2 \over 4 \pi^{3/2} \dot \phi},
\end{equation}
and the gravitational wave amplitude is given by
\begin{equation}
\label{eqtensoramplitude}
\left(h_{+,\times}\right)_{k = a H} = {H \over 2 \pi m_{\rm Pl}}
\end{equation}
for each of the two polarization modes for the gravitational wave. These are the amplitudes for
a single mode when its wavelength (which is changing with time due to expansion) is equal to the
horizon size. In the case of slow roll,
with $\dot\phi$ small and $H$ slowly varying, modes of different wavelengths will have 
approximately the same amplitudes, with slow variation as a function of scale. If we define the 
power spectrum as the variance per logarithmic interval,
\begin{equation}
\left({\delta \rho \over \rho}\right)^2 = \int{P_{\rm S}(k) d\log{k}},
\end{equation}
inflation generically predicts a power-law form for $P_{\rm S}(k)$, 
\begin{equation}
P_{\rm S}\left(k\right) \propto k^{n - 1},
\end{equation}
so that the {\em scale invariant spectrum}, one with equal amplitudes at horizon crossing,
is given by $n = 1$. The current observational best fit for the spectral index $n$ is
\cite{wangtegmarkzald} 
\begin{equation}
n = 0.91^{+ 0.15}_{-0.07}.
\end{equation}
The observations are in agreement with inflation's prediction of a nearly
(but not exactly) scale-invariant power spectrum, corresponding to a slowly rolling inflaton
field and a slowly varying Hubble parameter during inflation. One can also consider power
spectra which deviate from a power law, 
\begin{equation}
{d \log P_{\rm S}(k) \over d \log k} = n + {d n \over d \log{k}} + \cdots,
\end{equation}
but inflation predicts the variation in the spectral index $d n / d \log{k}$ to be small,
and we will not consider it further here. Similarly, the tensor fluctuation spectrum in 
inflationary models is a power-law,
\begin{equation}
P_{\rm T}\left(k\right) \propto k^{n_{\rm T}}.
\end{equation}
(Note the unfortunate convention that the scalar spectrum is defined as a power law with
index $n - 1$ while the tensor spectrum is defined as a power law with index $n_{\rm T}$,
so that the scale-invariant limit for tensors is $n_{\rm T} = 0$!)

It would appear, then, that we have four independent observable quantities to work with:
the amplitudes of the scalar and tensor power spectra at some fiducial scale $k_*$: 
\begin{eqnarray}
P_{\rm S} &\equiv& P_{\rm S}\left(k_*\right)\cr
P_{\rm T} &\equiv& P_{\rm T}\left(k_*\right),
\end{eqnarray}
and the spectral indices $n$,$n_{\rm T}$ of the power spectra. (Generally, the scale $k_*$ is
taken to be about the scale of the horizon size today, corresponding to the CMB quadrupole.)
In fact, at least within the
context of inflation driven by a single scalar field, not all of these parameters are
independent. This is because the tensor spectral index is just given by the equation of
state parameter $\epsilon$,
\begin{equation}
n_{\rm T} = - 2 \epsilon = - {m_{\rm Pl}^2 \over 8 \pi } \left[{V'\left(\phi\right) \over V\left(\phi\right)}\right]^2.
\end{equation}
However, from Eqs. (\ref{eqscalaramplitude}) and (\ref{eqtensoramplitude}), and from the
equation of motion in the slow-roll approximation (\ref{eqslowrollEOM}), we have a simple
expression for the ratio between the amplitudes of tensor and scalar fluctuations:
\begin{equation}
{P_{\rm T} \over P_{\rm S}} = 4 \pi  \left({\dot\phi \over m_{\rm Pl} H}\right)^2 = \epsilon = - n_{\rm T} / 2,
\end{equation}
so that the tensor/scalar ratio and the tensor spectral index are not independent parameters,
but are both determined by the equation of state during inflation, a relation known as the
{\em consistency condition} for slow-roll inflation.\footnote{In the case of multi-field 
inflation, this condition relaxes to an inequality, $P_{\rm T} / P_{\rm S} \geq -2 n_{\rm T}$
\cite{multifieldcons}.} 
It is conventional to define the tensor scalar ratio as the ratio $r$ of the contributions of
the modes to the CMB quadrupole, which adds roughly a factor of 10 \cite{turnerlambda}:
\begin{equation}
r \equiv {C_2^{\rm T} \over C_2^{\rm S}} \simeq 10 {P_{\rm T} \over P_{\rm S}} = 10 \epsilon.
\end{equation}
Similarly, the scalar spectral index can be expressed in terms of a second slow roll 
parameter $\eta$,
\begin{equation}
n - 1 = 4 \epsilon - 2 \eta,
\end{equation}
where $\eta$ depends on the second derivative of the potential,
\begin{equation}
\eta \equiv {m_{\rm Pl}^2 \over 8 \pi}\left[{V''\left(\phi\right) \over V\left(\phi\right)} - {1 \over 2} \left({V'\left(\phi\right) \over V\left(\phi\right)}\right)^2\right].
\end{equation}

So any simple inflation model gives us three independent parameters to describe the primordial
power spectrum: the amplitude of scalar fluctuations $A_{\rm S}$, the tensor/scalar ratio $r$,
and the scalar spectral index $n$. The important point is that these are {\em observable} 
parameters, and will allow us to make contact between the physics of very high energies and
the world of observational cosmology, in particular the cosmic microwave background. In the
next section, we will see in detail how to accomplish this for a simple model.

\subsection{A worked example}
\label{secinflationexample}

It will be useful to see how all this works in the context of a specific potential. We will choose
one of the simplest possible models, a massless field with a quartic self-coupling,
\begin{equation}
V\left(\phi\right) = \lambda \phi^4.
\end{equation}
This potential, simple as it is, has all the characteristics needed to support a successful
inflationary expansion. We will assume up front that the field is slowly rolling, so that
Eqs. (\ref{eqslowrollEOM},\ref{eqslowrollfriedmann}) describe the equations of motion
for the field, 
\begin{equation}
\label{eqphi4slowroll}
\dot\phi = -{V'\left(\phi\right) \over 3 H} = - \sqrt{ m_{\rm Pl}^2 \over 24 \pi} 
           {V'\left(\phi \right) \over \sqrt{V\left(\phi\right)}}.
\end{equation}
In order for inflation to occur, we must have negative pressure, $p < -(1/3)\rho$, which
is equivalent to the slow roll parameter $\epsilon$ being less than unity,
\begin{equation}
\epsilon = {m_{\rm Pl}^2 \over 16 \pi} \left[{V'\left(\phi\right) \over V\left(\phi\right)}\right]^2
         = {1 \over \pi} \left({m_{\rm Pl} \over \phi}\right)^2 < 1,
\end{equation}
so that inflation occurs when $\phi > \phi_{\rm e} = \left(m_{\rm Pl} / \sqrt{\pi}\right)$. Note
that the field is displaced a {\em long way} from the minimum of the potential at $\phi = 0$! 
This has been the source of some criticism of this type of model as a valid potential in an
effective field theory \cite{eftinflation}, but here we will accept this fact at the very least
as valid phenomenology. 

In this simple model, then, we have inflation happening when the field is rolling down the 
potential in a region far displaced from the minimum $\phi > m_{\rm Pl}$. Inflation ends
naturally at late time, when $\phi$ passes through $\phi_{\rm e} = m_{\rm Pl} / \sqrt{\pi}$. In 
order to solve the horizon and flatness problems, we must have at least a factor of $e^{55}$
expansion. The number of e-folds is given by Eq. (\ref{eqnumefoldsdef}),
\begin{equation}
N = \int{H dt} = \int{{H \over \dot\phi} d\phi}.
\end{equation}
It is convenient to choose the limits on the integral such that $N = 0$ at the {\em end} of
inflation, so that $N$ counts the number of e-folds until inflation ends and increases as we
go backward in time. Then, using the equation of motion for the field, we can show that
$N$ is just an integral over the slow-roll parameter $\epsilon$, and can be expressed as a
function of the field value $\phi$:
\begin{equation}
N\left(\phi\right) = {2 \sqrt{\pi}\over m_{\rm Pl}} \int_{\phi_{\rm e}}^{\phi} {d\phi' \over \sqrt{\epsilon\left(\phi'\right)}}.
\end{equation}
For our $\lambda \phi^4$ potential, the number of e-folds is 
\begin{eqnarray}
N\left(\phi\right) &=& {\pi \over m_{\rm Pl}^2} \left(\phi^2 - \phi_{\rm e}^2\right)\cr
                  &=& \pi \left({\phi \over m_{\rm Pl}}\right)^2 - 1.
\end{eqnarray}
Equivalently, we can define the field $\phi_N$ as the field value $N$ e-folds before the
end of inflation,
\begin{equation}
\phi_N \equiv m_{\rm Pl} \sqrt{N + 1 \over \pi}.
\end{equation}
We can now test our original assumption that the field is slowly rolling. It is simple to show
by differentiating Eq. (\ref{eqphi4slowroll}) that the acceleration of the field is given by
\begin{equation}
\ddot\phi = {2 m_{\rm Pl}^2 \over 3 \pi} \lambda \phi,
\end{equation}
which is indeed small relative to the derivative of the potential for $\phi \gg m_{\rm Pl}$:
\begin{equation}
{\ddot \phi \over V'\left(\phi\right)} = {1 \over 6 \pi}  \left({m_{\rm Pl} \over \phi}\right)^2.
\end{equation}
We see that slow roll is beginning to break down at the end of inflation, but is an excellent 
approximation for large $N$.

The rest is cookbook. We want to evaluate the power spectrum amplitude $P_{\rm S}$, 
tensor/scalar ratio $r$,
and scalar spectral index $n$ for fluctuations with scales comparable to the horizon size today,
which means fluctuations which crossed outside the horizon during inflation at about $N \simeq 55$.
Therefore, to calculate the amplitude $P_{\rm S}$, we evaluate
\begin{eqnarray}
P_{\rm S}^{1/2} &=& {H^2 \over 4 \pi^{3/2} \dot\phi}\cr
&=& 4 \sqrt{2 \over 3} {\left[V\left(\phi\right)\right]^{3/2} \over m_{\rm Pl}^3 V'\left(\phi\right)}
\end{eqnarray}
at $\phi = \phi_{55}$, or:
\begin{equation}
P_{\rm S}^{1/2} = \sqrt{2 \lambda \over 3} \left(\phi_{55} \over m_{\rm Pl}\right)^3 \sim 60 \lambda^{1/2}.
\end{equation}
But from the CMB, we know that the power spectrum amplitude is $P_{\rm S}^{1/2} \sim 10^{-5}$, 
so that means we must have a very tiny self-coupling for the field, $\lambda \sim 10^{-14}$. In order to 
sufficiently suppress the density fluctuation amplitude, the model must be extremely fine-tuned. 
This is a typical property of scalar field models of inflation. This also allows us to estimate
the energy scale of inflation,
\begin{equation}
E \sim V\left(\phi\right)^{1/4} = \lambda^{1/4} \phi_{55} \sim 10^{-3} m_{\rm PL} \sim 10^{16}\ {\rm GeV},
\end{equation}
or right about the scale of Grand Unification. This  interesting coincidence is
typical of most models of inflation. 

Finally, it is straightforward to calculate the tensor/scalar ratio and spectral index,
\begin{equation}
r = 10 \epsilon\left(\phi_{55}\right) = {10 \over 56} \simeq 0.18,
\end{equation}
and
\begin{equation}
n  = 1 - 4 \epsilon\left(\phi_{55}\right) + 2 \eta\left(\phi_{55}\right) = 1 - {3 \over 56} \simeq 0.95.
\end{equation}
The procedure for other potentials is similar: first, find the field value where inflation ends.
Then calculate the field value $55$ e-folds before the end of inflation and evaluate the expressions
for the observables at that field value. In this way we can match any given model of inflation
to its observational predictions. In the next section, we examine the predictions of different
types of models in light of current and future observational constraints, and find that it will
be possible with realistic measurements to distinguish between different models of inflation.

\subsection{Inflationary ``zoology'' and the CMB}
\label{secinflationzoology}

\begin{figure}[ht]
\vskip0.5cm
\centerline{\epsfig{file=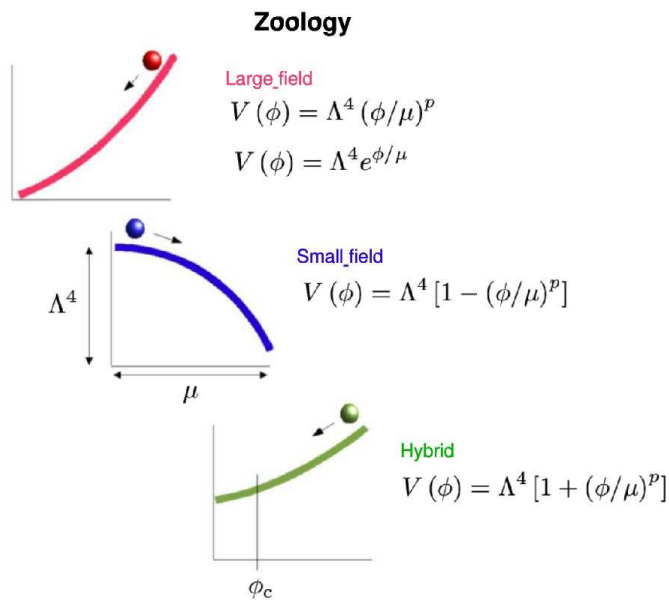, width=11.0cm}}
\caption{A ``zoology'' of inflation models, grouped into {\em large field}, {\em small field},
and {\em hybrid} potentials.}
\label{figzoology}
\end{figure}

\begin{figure}[ht]
\vskip0.5cm
\centerline{\epsfig{file=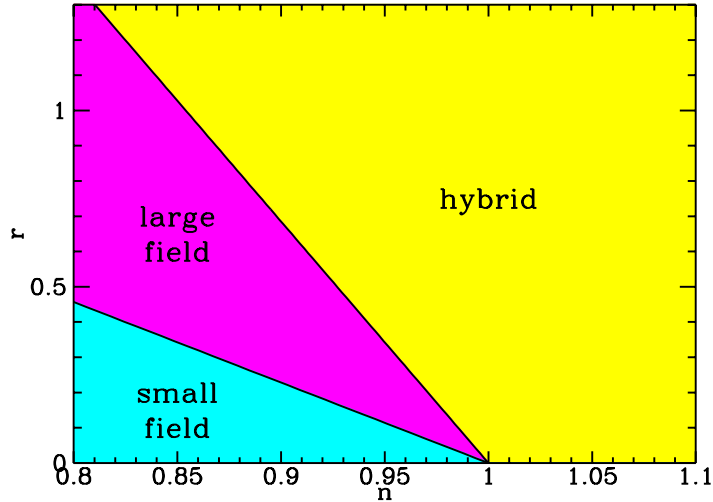, width=10.0cm}}
\caption{Classes of inflationary potentials plotted on the $(r,n)$ plane.}
\label{figzooplot}
\end{figure}

We have discussed in some detail the generation of fluctuations in inflation, in particular
the primordial tensor and scalar power spectra and the relevant parameters $A_{\rm S}$, $r$,
and $n$. We have also looked at one particular choice of potential $V(\phi) = \lambda \phi^4$ 
to drive inflation. In fact, pretty much {\em any} potential, with suitable fine-tuning of 
parameters,
will work to drive inflation in the early universe. We wish to come up with a classification
scheme for different kinds of scalar field potentials and study how we might find observational
constraints on one ``type'' of inflation versus another. Such a ``zoology'' of potentials
is simple to construct. Figure {\ref{figzoology}} shows three basic types of potentials: 
{\em large field}, where the field is displaced by $\Delta \phi \sim m_{\rm Pl}$ from a stable
minimum of a potential, {\em small field}, where the field is evolving away from an unstable
maximum of a potential, and {\em hybrid}, where the field is evolving toward a potential
minimum with nonzero vacuum energy. 
Large field models are perhaps the simplest types of potentials. These include potentials
such as a simple massive scalar field, $V(\phi) = m^2 \phi^2$, or fields with a quartic
self-coupling, $V(\phi) = \lambda \phi^4$. A general set of large field polynomial potentials
can be written in terms of a ``height'' $\Lambda$ and a ``width'' $\mu$ as:
\begin{equation}
V\left(\phi\right) = \Lambda^4 \left({\phi \over \mu}\right)^p,
\end{equation}
where the particular model is specified by choosing the exponent $p$. In addition, there is
another class of large field potentials which is useful, an exponential potential,
\begin{equation}
V\left(\phi\right) = \Lambda^4 \exp\left(\phi / \mu\right).
\end{equation}
Small field models are typical of spontaneous symmetry breaking, in which the field is
evolving away from an unstable maximum of the potential. In this case, we need not be concerned
with the form of the potential far from the maximum, since all the inflation takes place
when the field is very close to the top of the hill. As such, a generic potential of this
type can be expressed by the first term in a Taylor expansion about the maximum,
\begin{equation}
V\left(\phi\right) \simeq \Lambda^4 \left[1 - \left(\phi / \mu\right)^p\right],
\end{equation}
where the exponent $p$ differs from model to model. In the simplest case of spontaneous
symmetry breaking with no special symmetries, the leading term will be a mass term,
$p = 2$, and $V(\phi) = \Lambda^4 - m^2 \phi^2$. Higher order terms are also possible.
The third class of models we will call ``hybrid'', named after models first proposed
by Linde \cite{hybridinflation}. In these models, the field evolves toward the
minimum of the potential, but the  minimum has a nonzero vacuum energy, 
$V(\phi_{\rm min}) = \Lambda^4$. In such cases, inflation continues forever unless an
auxiliary field is added to bring an end to inflation at some point $\phi = \phi_{\rm c}$. 
Here we will treat the effect of this auxiliary field as an additional free parameter.
We will consider hybrid models in a similar fashion to large field and small field models,
\begin{equation}
V\left(\phi\right) = \Lambda^4 \left[1 + \left(\phi / \mu\right)^p\right].
\end{equation}
Note that potentials of all three types are parameterized in terms of a height $\Lambda^4$,
a width $\mu$, and an exponent $p$, with an additional parameter $\phi_{\rm c}$ specifying
the end of inflation in hybrid models. This classification of models may seem somewhat 
arbitrary.\footnote{The classification might also appear ill-defined, but it can be made 
more rigorous as a set of inequalities between slow roll parameters. Refs. 
\cite{dodelsonkolbkinney,kinney98} contain a more detailed discussion.} It is convenient,
however, because the different classes of models cover different regions of the $(r,n)$
plane with no overlap (Fig. \ref{figzooplot}).

\begin{figure}[ht]
\vskip0.5cm
\centerline{\epsfig{file=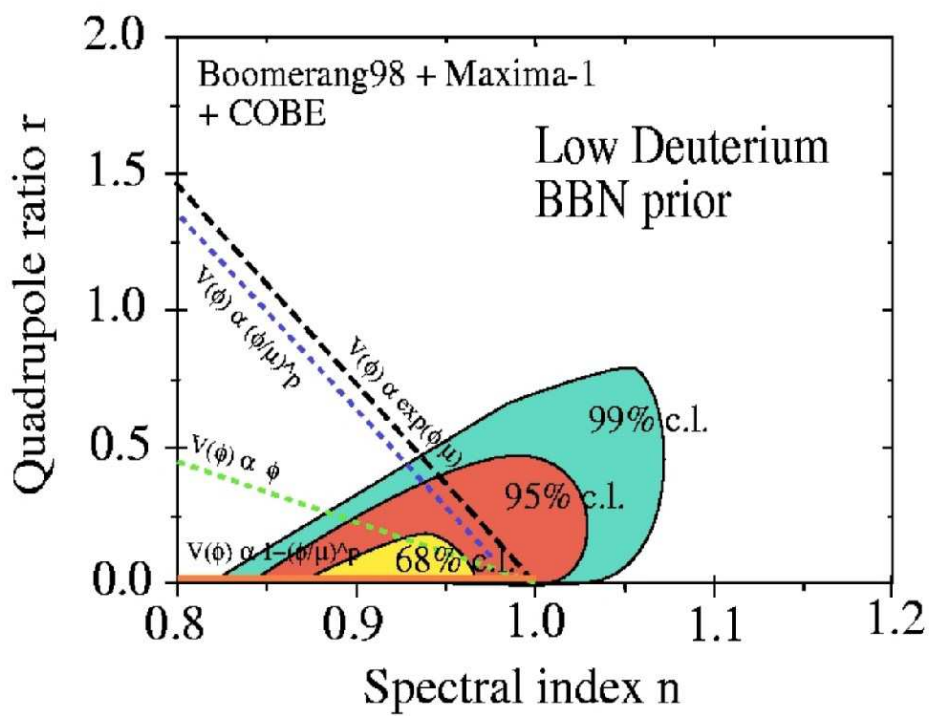, width=14.0cm}}
\caption{Error bars in the $r$, $n$ plane for the Boomerang and MAXIMA data sets 
\cite{kinneyriottomelchiorri}. The lines
on the plot show the predictions for various potentials. Similar contours for the most current
data can be found in Ref. \cite{wang02}.}
\label{figcurrentzooerrors}
\end{figure}

\begin{figure}[ht]
\vskip0.5cm
\centerline{\epsfig{file=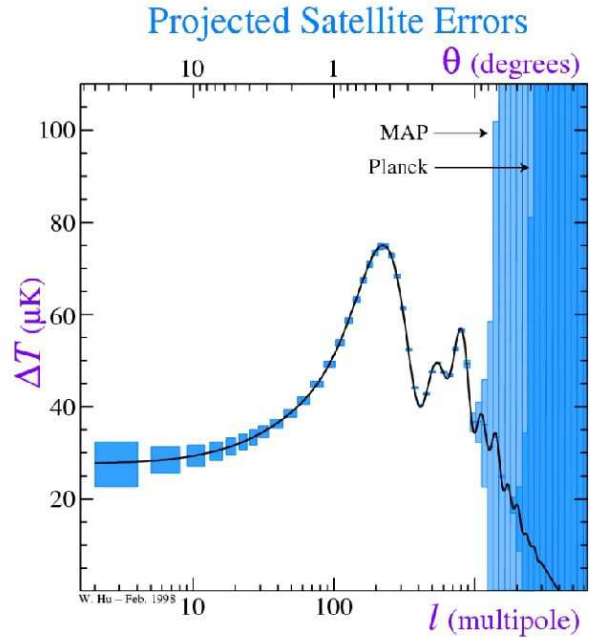, width=14.0cm}}
\caption{Expected errors in the $C_\ell$ spectrum for the MAP (light blue) and Planck (dark blue)
satellites. (Figure courtesy of Wayne Hu \cite{waynehuwebsite}.)}
\label{figMAPPlanckCl}
\end{figure}

\begin{figure}[ht]
\vskip0.5cm
\centerline{\epsfig{file=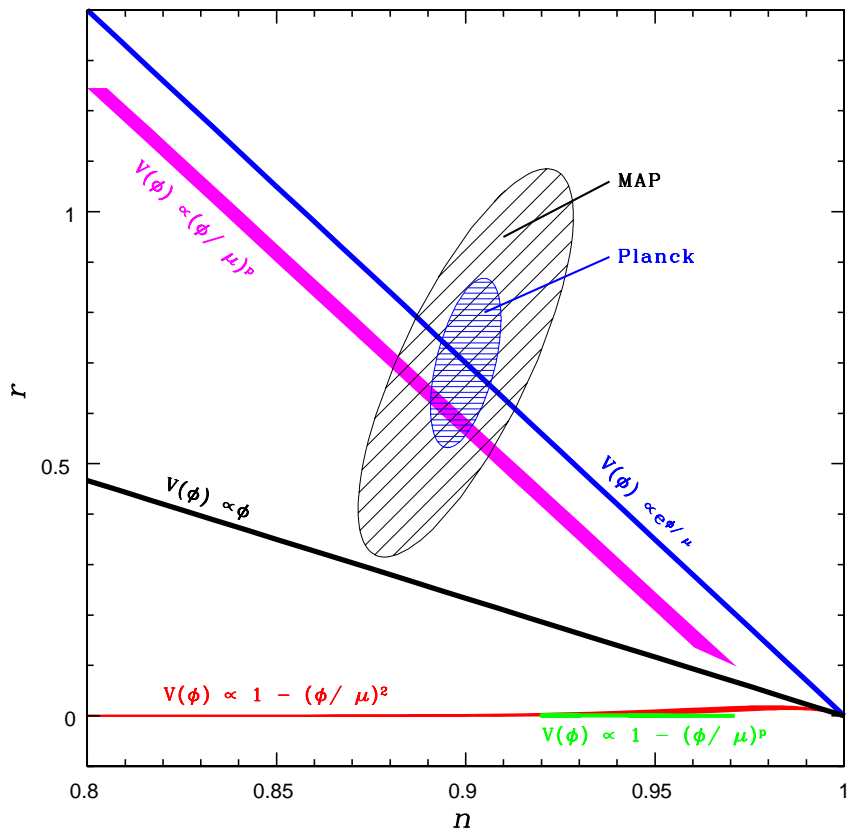, width=12.0cm}}
\caption{Error bars in the $r$, $n$ plane for MAP and Planck \cite{kinney98}. These ellipses show the
expected $2-\sigma$ errors. The lines on the plot show the predictions for various potentials. Note
that these are error bars based on {\em synthetic} data: the size of the error bars is meaningful,
but not their location on the plot. The best fit point for $r$ and $n$ from real data is likely
to be somewhere else on the plot.}
\label{figMAPPlanckzooerrors}
\end{figure}

How well can we constrain these parameters with observation? Figure \ref{figClflatopen} shows
the current observational constraints on the CMB multipole spectrum. As we saw in Sec. 
\ref{seccmb}, the shape of this spectrum depends on a large number of parameters, among
them the shape of the primordial power spectrum. However, uncertainties in other parameters
such as the baryon density or the redshift of reionization can confound our measurement of
the things we are interested in, namely $r$ and $n$. Figure \ref{figcurrentzooerrors} shows
likelihood contours for $r$ and $n$ based on current CMB data \cite{kinneyriottomelchiorri}.
Perhaps the most distinguishing feature of this plot is that the error bars are smaller than
the plot itself! The favored model is a model with negligible tensor fluctuations and a
slightly ``red'' spectrum, $n < 1$. Future measurements, in particular the MAP \cite{refMAP}
and Planck satellites \cite{refPlanck}, will provide much more accurate measurements of the
$C_\ell$ spectrum, and will allow correspondingly more precise determination of cosmological
parameters, including $r$ and $n$. (In fact, by the time this article sees print, the first
release of MAP data will have happened.) Figure \ref{figMAPPlanckCl} shows the expected
errors on the $C_\ell$ spectrum for MAP and Planck, and Fig. \ref{figMAPPlanckzooerrors} 
shows the corresponding error bars in the $(r,n)$ plane. Note especially that Planck will
make it possible to clearly distinguish between different models for inflation.
So all of this apparently esoteric theorizing about the extremely early universe is {\em not} 
idle speculation, but real science. Inflation makes a number of specific and observationally
testable predictions, most notably the generation of density and gravity-wave fluctuations
with a nearly (but not exactly) scale-invariant spectrum, and that these fluctuations are
Gaussian and adiabatic. Furthermore, feasible cosmological observations are capable of telling
apart different specific models for the inflationary epoch, thus providing us with real
information on physics near the expected scale of Grand Unification, far beyond the reach
of existing accelerators. In the next section,
we will stretch this idea even further. Instead of using cosmology to test the physics of 
inflation, we will discuss the more speculative idea that we might be able to use inflation
itself as a microscope with which to illuminate physics at the very highest energies, where
quantum gravity becomes relevant.

\section{Looking for signs of quantum gravity in inflation}
\label{sectransplanck}

We have seen that inflation is a powerful and predictive theory of the physics of the
very early universe. Unexplained properties of a standard FRW cosmology, namely the flatness
and homogeneity of the universe, are natural outcomes of an inflationary expansion. Furthermore,
inflation provides a mechanism for generating the tiny primordial density fluctuations that
seed the later formation of structure in the universe. Inflation makes definite predictions,
which can be tested by precision observation of fluctuations in the CMB, a program that is
already well underway. 

In this section, we will move beyond looking at inflation as a subject of experimental test
and discuss  some intriguing new ideas that indicate that inflation might be useful as a tool
to illuminate physics at extremely high energies, possibly up to the point where effects 
from quantum gravity become relevant. This idea is based on a simple observation about scales
in the universe. As we discussed in Sec. \ref{secintro}, quantum field theory extended to 
infinitely high energy scales gives nonsensical (i.e., divergent) results. We therefore 
expect the theory to break down at high energy, or equivalently at very short lengths. We
can estimate the length scale at which quantum mechanical effects from gravity become 
important by simple dimensional analysis. We define the Planck length $\ell_{\rm Pl}$
by an appropriate combination of fundamental constants as
\begin{equation}
\ell_{\rm Pl} \sim \sqrt{\hbar G \over c^3} \sim 10^{-35} m.
\end{equation}
For processes probing length scales shorter than $\ell_{\rm Pl}$, such as quantum modes with 
wavelengths $\lambda < \ell_{\rm Pl}$, we expect some sort of new physics to be important.
There are a number of ideas for what that new physics might be, for example string theory
or noncommutative geometry or discrete spacetime, but physics at the Planck scale is currently
not well understood. It is unlikely that particle accelerators will provide insight into
such high energy scales, since quantum modes with wavelengths less than $\ell_{\rm Pl}$ will
be characterized by energies of order $10^{19}\ {\rm GeV}$ or so, and current particle 
accelerators operate at energies around $10^{3}\ {\rm GeV}$.\footnote{ This might not be
so in ``braneworld'' scenarios where the energy scale of quantum gravity can be much lower
\cite{braneworld}.}
However, we note an interesting
fact, namely that the ratio between the current horizon size of the universe and the Planck
length is about
\begin{equation}
{d_{\rm H} \over l_{\rm Pl}} \sim 10^{60},
\end{equation}
or, on a log scale,
\begin{equation}
\ln\left({d_{\rm H} \over l_{\rm Pl}}\right) \sim 140.
\end{equation}
This is a big number, but we recall our earlier discussion of the flatness and horizon problems
and note that inflation, in order to adequately explain the flatness and homogeneity of the
universe, requires the scale factor to increase by {\em at least} a factor of $e^{55}$. Typical
models of inflation predict much more expansion, $e^{1000}$ or more. We remember that the
wavelength of quantum modes created during the inflationary expansion, such as those responsible
for density and gravitational-wave fluctuations, have wavelengths which redshift proportional
to the scale factor, so that so that the wavelength $\lambda_i$ of a mode at early times 
can be given in terms of its wavelength $\lambda_0$ today by
\begin{equation}
\lambda_i \ll e^{-N} \lambda_0.
\end{equation}
This means that if inflation lasts for more than about $N \sim 140$ e-folds, fluctuations of 
order the size of the universe
today were smaller than the Planck length during inflation! This suggests the possibility that
Plank-scale physics might have been important for the generation of quantum modes in inflation.
The effects of such physics might be imprinted in the pattern of cosmological fluctuations
we see in the CMB and large-scale structure today. In what follows, we will look at the
generation of quantum fluctuations in inflation in detail, and estimate how large the effect
of quantum gravity might be on the primordial power spectrum.

In Sec. \ref{secqftbasics} we saw that the state space for a quantum field theory was a set
of states $\left\vert n({\bf k}_1),\ldots,n({\bf k}_i)\right\rangle$ representing
the number of particles with momenta ${\bf k}_1,\ldots,{\bf k}_i$. The creation and annihilation
operators $\hat a^{\dagger}_{\bf k}$ and ${\hat a_{\bf k}}$ act on these states by adding or subtracting
a particle from the state:
\begin{eqnarray}
\hat a^{\dagger}_{\bf k} \left\vert n({\bf k})\right\rangle &&= \sqrt{n + 1} \left\vert n({\bf k}) + 1\right\rangle\cr
{\hat a_{\bf k}} \left\vert n({\bf k})\right\rangle &&= \sqrt{n} \left\vert n({\bf k}) - 1\right\rangle.
\end{eqnarray}
The ground state, or vacuum state of the space is just the zero particle state,
\begin{equation}
{\hat a_{\bf k}} \left\vert 0 \right\rangle = 0.
\end{equation}
Note in particular that the vacuum state $\left\vert 0 \right\rangle$ is {\em not} equivalent
to zero. The vacuum is not nothing:
\begin{equation}
\left\vert 0 \right\rangle \neq 0.
\end{equation}
To construct a quantum field, we look at the familiar classical wave equation for a scalar field,
\begin{equation}
\label{eqminkowskiwaveequation}
{\partial^2 \phi \over \partial t^2} - \nabla^2 \phi = 0.
\end{equation}
To solve this equation, we decompose into Fourier modes $u_{\rm k}$,
\begin{equation}
\label{eqfourierexpansion}
\phi =  \int{d^3 k \left[a_{\bf k} u_{\bf k}(t) e^{i {\bf k}\cdot{\bf x}} + a^*_{\bf k} u^*_{\bf k}(t) e^{- i {\bf k}\cdot{\bf x}}\right]},
\end{equation}
where the mode functions $u_{\bf k}(t)$ satisfy the ordinary differential equations
\begin{equation}
\label{eqminkowskimode}
\ddot u_{\bf k} + k^2 u_{\bf k} = 0.
\end{equation}
This is a classical wave equation with a classical solution, and the Fourier coefficients
$a_{\rm k}$ are just complex numbers. The solution for the mode function is 
\begin{equation}
\label{eqminkowskimodefunction}
u_{\bf k} \propto e^{-i \omega_k t},
\end{equation}
where $\omega_k$ satisfies the dispersion relation
\begin{equation}
\omega_k^2 - {\bf k}^2 = 0.
\end{equation}
To turn this into a quantum field, we identify the
Fourier coefficients with creation and annihilation operators
\begin{equation}
a_{\bf k} \rightarrow \hat a_{\bf k},\ a^*_{\bf k} \rightarrow \hat a^{\dagger}_{\bf k},
\end{equation}
and enforce the commutation relations
\begin{equation}
\left[\hat a_{\bf k}, \hat a^{\dagger}_{\bf k'}\right] = \delta^3\left({\bf k} - {\bf k'}\right).
\end{equation}
This is the standard quantization of a scalar field in 
Minkowski space, which should be familiar. But what probably isn't familiar is that this solution
has an interesting symmetry. Suppose we define a new mode function $u_{\bf k}$ which 
is a rotation of the solution (\ref{eqminkowskimodefunction}):
\begin{equation}
\label{eqrotatedmodefunction}
u_{\bf k} = A(k) e^{-i \omega t + i {\bf k} \cdot {\bf x}} + B(k) e^{i \omega t - i {\bf k} \cdot {\bf x}}.
\end{equation}
This is {\em also} a perfectly valid solution to the original wave equation 
(\ref{eqminkowskiwaveequation}), since it is just a superposition of the Fourier modes. But we 
can then re-write the quantum
field in terms of our original Fourier modes and new {\em operators} $\hat b_{\bf k}$ and 
$\hat b^{\dagger}_{\bf k}$ and the original Fourier modes $e^{i {\bf k} \cdot {\bf x}}$ as:
\begin{equation}
\phi =  \int{d^3 k \left[{\hat b_{\bf k}} e^{-i \omega t + i {\bf k}\cdot{\bf x}} + \hat b^{\dagger}_{\bf k}  e^{+ i \omega t - i {\bf k}\cdot{\bf x}}\right]},
\end{equation}
where the new operators $\hat b_{\bf k}$ are given in terms of the old operators ${\hat a_{\bf k}}$ by
\begin{equation}
\hat b_{\bf k} = A(k) \hat a_{\bf k} + B^*(k) \hat a^{\dagger}_{\bf k}.
\end{equation}
This is completely equivalent to our original solution (\ref{eqfourierexpansion}) as long as the
new operators satisfy the same commutation relation as the original operators,
\begin{equation}
\left[\hat b_{\bf k}, \hat b^{\dagger}_{\bf k'}\right] = \delta^3\left({\bf k} - {\bf k'}\right).
\end{equation}
This can be shown to place a condition on the coefficients $A$ and $B$,
\begin{equation}
\left\vert A\right\vert^2 - \left\vert B\right\vert^2 = 1.
\end{equation}
Otherwise, we are free to choose $A$ and $B$ as we please. 

This is just a standard property of linear differential equations: any linear combination of
solutions is itself a solution. But what does it mean physically? In one case, we have an
annihilation operator ${\hat a_{\bf k}}$ which gives zero when acting on a particular state which
we call the vacuum state:
\begin{equation}
{\hat a_{\bf k}} \left\vert 0_a \right\rangle = 0.
\end{equation}
Similarly, our rotated operator $\hat b_{\bf k}$ gives zero when acting on some state
\begin{equation}
\hat b_{\bf k} \left\vert 0_b\right\rangle = 0.
\end{equation}
The point is that the two ``vacuum'' states are not the same
\begin{equation}
\left\vert 0_a \right\rangle \neq \left\vert 0_b\right\rangle.
\end{equation}
From this point of view, we can define any state we wish to be the ``vacuum'' and build a
completely consistent quantum field theory based on this assumption. From another equally 
valid point of view this state will contain particles. How do we tell which is the 
{\em physical} vacuum state? To define the real vacuum, we have to consider the spacetime
the field is living in. For example, in regular special relativistic quantum field theory,
the ``true'' vacuum is the zero-particle state as seen by an inertial observer. Another
more formal way to state this is that we require the vacuum to be Lorentz symmetric. This
fixes our choice of vacuum $\left\vert 0\right\rangle$ and defines unambiguously our set
of creation and annihilation operators $\hat a$ and $\hat a^{\dagger}$. A consequence of this is that
an {\em accelerated} observer in the Minkowski vacuum will think that the space is full
of particles, a phenomenon known as the Unruh effect \cite{unruheffect}. The zero-particle
state for an accelerated observer is different than for an inertial observer. 

So far we have been working within the context of field theory in special relativity. What
about in an expanding universe? The generalization to a curved spacetime is straightforward,
if a bit mysterious. We will replace the metric for special relativity with a Robertson-Walker
metric,
\begin{equation}
ds^2 = dt^2 - d{\bf x}^2 \rightarrow ds^2 = a^2\left(\tau\right) \left(d\tau^2 - d{\bf x}^2\right).
\end{equation}
Note that we have written the Robertson-Walker metric in terms of conformal time $d \tau = dt / a$.
This is convenient for doing field theory, because the new spacetime is just a Minkowski
space with a time-dependent conformal factor out front. In fact we define the physical vacuum
in a similar way to how we did it for special relativity: the vacuum is the zero-particle
state as seen by a {\em geodesic} observer, that is, one in free-fall in the expanding
space. This is referred to as the {\em Bunch-Davies} vacuum. 

Now we write down the wave equation for a free field, the equivalent of 
Eq. (\ref{eqminkowskiwaveequation}) in a Robertson-Walker space. This is the usual
equation with a new term that comes from the expansion of the universe:
\begin{equation}
{\partial^2 \phi \over \partial \tau^2} + 2 \left({a'(\tau) \over a(\tau)}\right) {\partial \phi \over \partial \tau} - \nabla^2 \phi = 0.
\end{equation}
Note that the time derivatives are with respect to the conformal time $\tau$, not the coordinate
time $t$. As in the Minkowski case, we Fourier expand the field, but with an extra factor 
of $a(\tau)$ in the integral:
\begin{equation}
\label{eqFRWfourier}
\phi =  \int{{d^3 k \over a(\tau)} \left[{\hat a_{\bf k}} u_{\bf k}(\tau) e^{i {\bf k}\cdot{\bf x}} + \hat a^{\dagger}_{\bf k} u^*_{\bf k}(\tau) e^{- i {\bf k}\cdot{\bf x}}\right]}.
\end{equation}
Here $k$ is a comoving wavenumber (or, equivalently, momentum), which stays constant as the
mode redshifts with the expansion $\lambda \propto a$, so that
\begin{equation}
k_{\rm physical} = k / a.
\end{equation}
Writing things this way, in terms of conformal time and comoving wavenumber, makes the equation
of motion for the mode $u_{\bf k}(\tau)$ very similar to the mode equation (\ref{eqminkowskimode})
in Minkowski space:
\begin{equation}
u''_{\bf k} + \left(k^2 - {a'' \over a}\right) u_{\bf k} = 0,
\end{equation}
where a prime denotes a derivative with respect to conformal time. All of the effect of the
expansion is in the $a'' / a$ term. (Be careful not to confuse the scale factor $a(\tau)$ with
the creation/annihilation operators ${\hat a_{\bf k}}$ and $\hat a^{\dagger}_{\bf k}$!)

\begin{figure}[ht]
\vskip0.5cm
\centerline{\epsfig{file=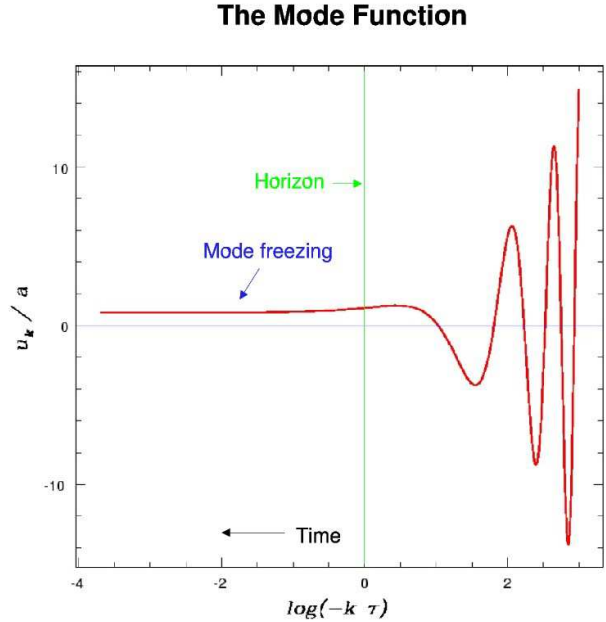, width=12.0cm}}
\caption{The mode function $u_{\bf k} / a$ as a function of 
$d_{\rm H} / \lambda = k / (a H) = -k \tau$.
At short wavelengths, $k \gg a H$, the  mode is oscillatory, but ``freezes out'' to a nonzero
value at long wavelengths, $k \ll a H$.}
\label{figmodefunction}
\end{figure}

This equation is easy to solve. First consider the short wavelength limit, that is large
wavenumber $k$. For $k^2 \gg a''/a$, the mode equation is just what we had for Minkowski
space
\begin{equation}
u''_{\bf k} + k^2 u_{\bf k} \simeq 0,
\end{equation}
except that we are now working with comoving momentum and conformal time, so the space is
only quasi-Minkowski. The general solution for the mode is
\begin{equation}
\label{eqgeneralRWmode}
u_{\bf k} = A(k) e^{-i k \tau} + B(k) e^{+ i k \tau}.
\end{equation}
Here is where the definition of the vacuum comes in. Selecting the Bunch-Davies vacuum is
equivalent to setting $A = 1$ and $B = 0$, so that the annihilation operator is multiplied
by $e^{-i k \tau}$ and not some linear combination of positive and negative frequencies. This
is the exact analog of Eq. (\ref{eqrotatedmodefunction}). So the mode function corresponding to
the zero-particle state for an observer in free fall is
\begin{equation}
\label{eqbunchdaviesstate}
u_{\bf k} \propto e^{-i k \tau}.
\end{equation}
What about the long wavelength limit, $k^2 \ll a''/a$? The mode equation becomes trivial:
\begin{equation}
u''_{\bf k} - {a'' \over a} u_{\bf k} = 0,
\end{equation}
with solution
\begin{equation}
\label{eqlongwavelength}
\left\vert {u_{\bf k} \over a}\right\vert = {\rm const.}
\end{equation}
The mode is said to be {\em frozen} at long wavelengths, since the oscillatory behavior is
damped. This is precisely the origin of the density and gravity-wave fluctuations in inflation. 
Modes at short wavelengths are rapidly redshifted by the inflationary expansion so that the 
wavelength of the mode is larger than the horizon size, Eq. (\ref{eqlongwavelength}). We
can plot the mode as a function of its physical wavelength $\lambda = k / a$ divided by
the horizon size $d_{\rm H} = H^{-1}$ (Fig. \ref{figmodefunction}), and find that at
long wavelengths, the mode freezes out to a nonzero value.
The power spectrum of fluctuations is just given by the two-point correlation function of the
field, 
\begin{equation}
P(k) \propto \left\langle \phi^2 \right\rangle_{k >> a H} \propto \left\vert u_{\bf k} \over a\right\vert^2 \neq 0.
\end{equation}
This means that we have produced {\em classical} perturbations at long wavelength from quantum
fluctuations at short wavelength. 

What does any of this have to do with quantum gravity? Remember that we have seen that for
an inflationary period that lasts longer than 140 e-folds or so, the fluctuations we see
with wavelengths comparable to the horizon size today started out with wavelengths shorter
than the Planck length $\ell_{\rm Pl} \sim 10^{-35}\ {\rm cm}$ during inflation. For a mode with
a wavelength that short, do we really know how to select the ``vacuum'' state, which we
have assumed is given by Eq. (\ref{eqbunchdaviesstate})? Not necessarily. We do know that once
the mode redshifts to a wavelength greater than $\ell_{\rm Pl}$, it must be of the form
(\ref{eqgeneralRWmode}), but we know longer know for certain how to select the values
of the constants $A(k)$ and $B(k)$. What we have done is mapped the effect of quantum gravity onto
a boundary condition for the mode function $u_{\bf k}$. In principle, $A(k)$ and $B(k)$ could
be anything! If we allow $A$ and $B$ to remain arbitrary, it is simple to calculate the
change in the two-point correlation function at long wavelength,
\begin{equation}
P(k) \rightarrow \left\vert A(k) + B(k)\right\vert^2 P_{\rm B-D}(k),
\end{equation}
where the subscript ${\rm B-D}$ indicates the value for the case of the ``standard'' Bunch-Davies
vacuum, which corresponds to the choice $A = 1$, $B = 0$. So the power spectrum of gravity-wave 
and density fluctuations is sensitive to how we
choose the vacuum state at distances shorter than the Planck scale, and is in principle 
sensitive to quantum gravity.

\begin{figure}[ht]
\vskip0.5cm
\centerline{\epsfig{file=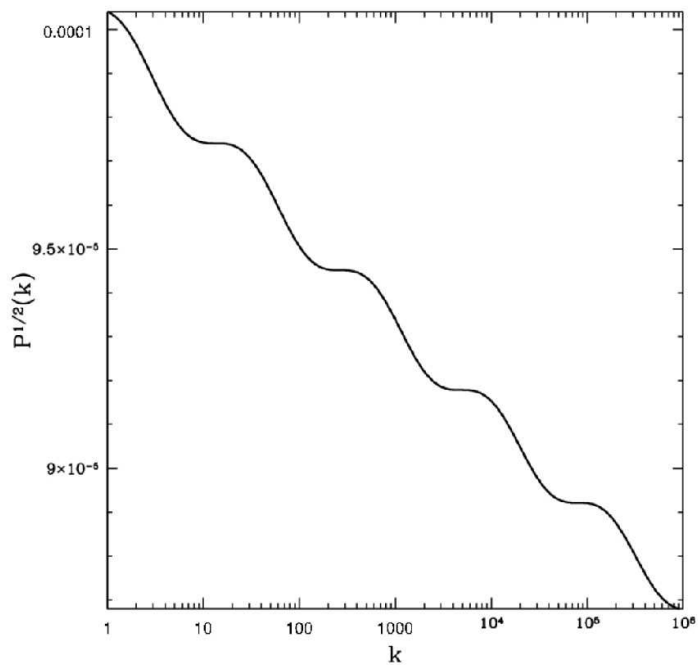, width=12.0cm}}
\caption{Modulation of the power spectrum of primordial fluctuations for a rotation 
$B \sim H / m_{\rm Pl}$. }
\label{figpsmodulation}
\end{figure}

While in principle $A(k)$ and $B(k)$ are arbitrary, a great deal of recent work has been done
implementing this idea within the context of reasonable toy models of the physics of short
distances. There is some disagreement in the literature with regard to how big the parameter
$B$ can reasonably be. As one might expect on dimensional grounds, the size of the rotation
is determined by the dimensionless ratio of the Planck length to the horizon size, so it is
expected to be small
\begin{equation}
B \sim \left({l_{\rm Pl} \over d_{\rm H}}\right)^p \sim \left({H \over m_{\rm Pl}}\right)^p\ll 1.
\end{equation}
Here we have introduced a power $p$ on the ratio, which varies depending on which model of
short-distance physics you choose. Several groups have shown an effect linear in the ratio,
$p = 1$. Fig. \ref{figpsmodulation} shows the modulation of the power spectrum calculated
in the context of one simple model \cite{transplanck1}.
Others have argued that this is too optimistic, and that a more realistic estimate
is $p = 2$ \cite{transplanck2} or even smaller \cite{transplanck3}. The difference is important: 
if $p = 1$, the modulation of the power spectrum
can be as large as a percent or so, a potentially observable value \cite{transplanckobs}. Take 
$p = 2$ and the modulation drops to a hundredth of a percent, far too small to see. Nonetheless, 
it is almost certainly worth looking for!

\section{Conclusion}

We have come a long way in four lectures, from Einstein's misbegotten introduction of the
cosmological constant at the beginning of the last century to its triumphant return today.
Einstein's blunder is now seen as the key to understanding the very beginning of the universe,
as represented by the theory of inflation, as well as the universe today, dominated by
the mysterious dark energy that makes up more than two thirds of the entire mass of the cosmos.
I have tried to convince you of two things: first, that the study of the early universe 
{\em is} particle physics in a very real sense, and second that apparently
exotic theories of the early universe such as inflation (and perhaps even elements of string 
theory or some other variant of quantum gravity) are predictive and testable.
 It is a difficult business, to be sure, compared to the clean physics at, say,
an ${\rm e}^{\pm}$ collider, but what we learn about fundamental theory from cosmology is
in many ways complementary to the lessons learned from more traditional particle physics. 

\section*{Acknowledgments}

I would like to give my warmest thanks to Harrison Prosper for organizing a wonderful summer
school, and to the bright and motivated students who made lecturing there a rare treat. I thank
Richard Easther and Brian Greene for helpful comments. ISCAP gratefully 
acknowledges the generous support of the Ohrstrom foundation.

\end{document}